\documentclass[12pt, draftclsnofoot, onecolumn]{IEEEtran}
\usepackage{filecontents}
\usepackage{amsthm}
\usepackage[table,xcdraw]{xcolor}
\newtheorem{theorem}{\textbf{\text{Theorem}}}
\newtheorem{corollary}{Corollary}
\newtheorem{lemma}{Lemma}

\newtheorem{claim}{Claim}
\usepackage{epsfig}
\usepackage{times}
\usepackage{verbatim}
\usepackage{enumerate}
\usepackage{multicol,afterpage,wrapfig}
\usepackage{cite}
\usepackage{graphicx}
\usepackage{algorithm}
\usepackage{algorithmic}
\usepackage{ifmtarg}
\usepackage{amssymb}
\usepackage{amsmath}
\usepackage{color}
\usepackage{bbm}
\usepackage{epstopdf}
\usepackage{flushend}
\usepackage{subcaption}
 \newcommand*{\bS}{\tilde{\sigma}^2}
\newtheorem{definition}{\textbf{\text{Definition}}}
\newtheorem{assumption}{Assumption}
 \usepackage{multirow}
 \usepackage{stackengine}
 \newcommand\barbelow[1]{\stackunder[1.2pt]{$#1$}{\rule{.8ex}{.075ex}}}
 
\makeatletter
\newcommand*{\rom}[1]{\expandafter\@slowromancap\romannumeral #1@}
\makeatother

\begin{document}
	\title{Stability and Metastability of Traffic Dynamics in Uplink Random Access Networks}
	\author{
		\IEEEauthorblockN{\large  Ahmad AlAmmouri, Jeffrey G. Andrews, and Fran\c cois Baccelli}
		\thanks{The authors are with the Wireless Networking and Communications Group (WNCG), The University of Texas at Austin, Austin, TX 78712 USA. (Email: \{alammouri@utexas.edu, jandrews@ece.utexas.edu, francois.baccelli@austin.utexas.edu\}). Last revised \today.}\thanks{ This work was supported 
in part by the National Science Foundation under Grant NSF-CCF-1514275
and in part by the Simons Foundation under Grant 197982. Part of this work was presented in \cite{Stability_AlAmmouri19Conf}.}}
	
	\maketitle
	\begin{abstract}
		We characterize the stability, metastability, and the stationary regime of traffic dynamics in a single-cell uplink wireless system. The traffic is represented in terms of spatial birth-death processes, in which users arrive as a Poisson point process in time and space, each with a file to transmit to the base station. The service rate of each user is based on its signal to interference plus noise ratio, where the interference is from other active users in the cell. Once the file is fully transmitted, the user leaves the cell. We derive the necessary and sufficient condition for network stability, which is independent of the specific bounded path loss function.  A novel observation is that for a certain range of arrival rates, the network appears stable for a possibly long time, and then suddenly exhibits instability. This property, which is known in statistical physics but rarely observed in wireless communication, is called {\it metastability}. Finally, we propose two heuristic characterizations based on mean-field interpretation,  of the network steady-state regime when it exists. The first-order approximation is very simple to compute, but loose in some regimes, whereas the second-order approximation is more sophisticated but tight for the whole range of arrival rates.     
	\end{abstract}

\section{Introduction}
Random access uplink networks have assumed renewed importance given the current and future expected growth of Internet of Things (IoT) devices and connections, which are expected to dwarf human-operated devices in the coming decade \cite{The_Atzori10}.  Many IoT use cases are distinguished from the now-dominant data and video traffic by the massive numbers of devices which each have sporadic traffic to send.  In such a scenario, devices enter the network without warning, wish to transmit some data quickly and without going through a lengthy acquisition and scheduling process, and then go back to sleep.  The stability, latency, and scalability of such a random access scenario -- despite considerable study, as explained below -- is largely unknown, and very challenging to analyze.  This paper takes a step forward in this direction by studying the dynamics of a single-cell uplink wireless system, along with its stability. Our approach and analysis rely on tools from  mean-field theory along with queuing theory, and allows us to derive the exact necessary and sufficient stability condition of the system, along with some simple heuristics to describe the stationary regime when it exists.
 
\subsection{History and Motivation: Wired Random Access Networks}

The history of analyzing dynamics in communication networks goes back to  wired data networks, where multiple nodes share a common wire to a common destination \cite{Data_Bertsekas92}. A benchmark random access protocol for these networks that has been extensively analyzed in the past three decades is slotted Aloha, where $N$ nodes share the same resource to the destination, each has a packet arrival rate of $\lambda_i$, and node $i$ transmits at the beginning of each time slot with a probability $p_i$ if its queue is not empty. If more than one node transmit in the same time slot, then a collision is declared, and the packets are queued back at their sources. Although the slotted Aloha protocol is simple, its \emph{stability region} for general $N$ -- which is the set of arrival rates $\{\lambda_i, \forall i \in [1, \cdots N]\}$ that leads to stable queues -- is a long-standing open problem \cite{Ergodicity_Tsybakov79}.  In some special cases, the stability region is known.  For example, if the arrivals follow a Bernoulli process, then the exact stability region is known for the cases of $N=2$ \cite{Ergodicity_Tsybakov79}, $N=3$ \cite{Stability_Szpankowski94}, and for a few cases that require specific ratios between the arrival rates and the transmission probabilities \cite{Bonald_Wireless04}. Otherwise, we only have approximations and bounds on the stability region \cite{On_Rao88,Stability_Borst08,Asymptotic_Bordenave12}.

This problem is challenging because of the interactions between queues, such that the status of one queue depends on the status of the other queues and their service rates, commonly referred to as {\it interacting queues problems} \cite{On_Rao88}. One approach to analyze these problems is through mean-field limits. Briefly, a mean-field limit is a mathematical tool that allows a varying environment to be abstracted using its empirical average state. In other words, assume we have $N$ queues that interact with each other in such a way that the state of a queue and its service rate depend on the current state of the environment, namely the state of the other queues. In the mean-field limit, each queue observes the empirical average of the state of the environment (the state of the other queues). Hence, its evolves independently of the current state of the other queues. This results in an isolation of the queue from the current state of the environment, but still approximately captures the effect of the environment through its empirical average, which could be spatial or temporal. In many cases, the mean-field limit was proven to be  asymptotically exact, when the number of queues tends to infinity.  %Such an approach decorrelate the status of the queues from each other.  
We refer the reader to \cite{A_Bordenave07} for a tutorial on mean-field analysis. 

 Communication networks are among the many domains where mean-field limits have been extensively used. For example, the approximation for the stability condition of the Aloha protocol in \cite{Asymptotic_Bordenave12} is based on mean-field analysis, and the authors proved that this approximation is asymptotically exact for large $N$. Another interesting case is the same Aloha protocol, but with the nodes employing an exponential back-off mechanism after a transmission failure before reaccessing the resource. This mechanism is implemented in the Ethernet protocol.  It was proven in \cite{Ultimate_Aldous87} that when $N\rightarrow\infty$, this network is ultimately unstable regardless of the arrivals rates. More specifically, the network is {\it metastable} for all arrival rates.
 
In general terms, metastability is a property of some stochastic dynamical systems having single or multiple global stable solutions, and possibly some local ones, where the system stays in the neighbourhood of one of these stable points for a very long time, and then, due to an infrequent  large random fluctuation, switches to another stable point. For example, in \cite{Metastability_Antunes06,Stochastic_Antunes08,Toward_Genin08} and in the Ising model in statistical physics \cite{liggett2013stochastic}, the system has two stable regimes, but it stays for a long time in one of them before switching to the other one, due to some rare random event. This migration time typically grows {\it exponentially} with the network size. In some special cases, the system has a unique globally stable point and other locally stable points \cite{Multi_Vvedenskaya07}. Hence, as long as the system is operating in the neighbourhood of these locally stable points, it looks to the observer as if it had reached its global stable point, since it can remain there for a very long time. However, given enough time, the system eventually converges to its true globally stable regime.
 
A special case of the latter instance of metastability is when the global stable regime is an absorbing state, i.e., the system does not come back from this state as in \cite{Metastability_Baccelli17}, in which the absorbing state is at $\infty$, and in the SIS model in \cite{liggett2013stochastic}, in which the absorbing state is at $0$. In these cases, the system can operate around its locally stable points for a very long time, before shifting to the absorbing state and staying there forever. For example, in \cite{Metastability_Baccelli17} and in the slotted Aloha case we mentioned, the network eventually departs from its locally stable regime to the divergent one. Hence, it might look to the observer as being stable for a very long time, despite being ultimately unstable in the long-run.

 Note that the intuition behind the metastability of slotted Aloha with exponential back-off mechanism was already known before, but it was proven for the first time in \cite{Ultimate_Aldous87}. The mean-field limit gives an indication  for this case that the system might have two equilibrium points. As long as the system operates around the first one, it acts as if it were stable. However, if the system is pushed to operate beyond the second equilibrium point, possibly due to a surge in packet arrivals, it becomes unstable. To the best of our knowledge, there is no general theorem proving that if the mean-field analysis results in two equilibrium points, then the system is metastable. Rather, it has to be studied case-by-case. For further use of mean-field analysis  to study metastability in communications networks, see  \cite{Metastability_Antunes06,Stochastic_Antunes08,Toward_Genin08}.

\subsection{Related Work: Wireless Networks}

So far we have focused on wired networks. By moving to a wireless setting, the network geometry along with the service rate function make these interacting particles (queues) problems even more challenging, but also more interesting. Specifically, the locations of the users with respect to (w.r.t.) their serving and interfering BSs determine their received signal quality and thus their service rates. Hence, the collision model used in wired networks is not directly applicable to cellular networks, since users can adapt their transmit rates, usually based on the measured signal to interference plus noise ratio (SINR). 

In wireless settings, the majority of works in the literature are traffic-agnostic -- e.g., all nodes transmit all the time, which is also known as the {\it full-buffer} model -- and the geometry of the network can be accounted for using tools from stochastic geometry \cite{Stochastic_Baccelli10_2,A_Andrews11,Stochastic_Haenggi12}. The relatively small literature that deals with wireless network traffic can be divided into four categories based on the network model and service rate function: ($i$) an ad hoc network with a fixed rate function \cite{On_Zhong16,On_Chisci17}, ($ii$) an ad hoc network with an adaptive rate function \cite{Spatial_Sankararaman17}, ($iii$) a cellular network with a fixed rate function \cite{Heterogeneous_Zhong17,Spatiotemporal_Gharbieh17, Spatiotemporal_Gharbieh18,Self_Gharbieh19}, ($iv$) and a cellular network with an adaptive rate function \cite{Performance_Blaszczyszyn15}. For the fixed rate function, the nodes transmit with a fixed rate, and the transmission is successful if and only if the received SINR is higher than a predefined threshold. In the adaptive rate function, the nodes adapt their transmission rates to the SINR, e.g., as $\log_2(1+{\rm SINR})$.

In the first category, the authors in \cite{On_Zhong16,On_Chisci17} derived approximations for the network stability region. In \cite{Spatial_Sankararaman17}, an adaptive rate function was considered and the exact stability region was found along with an approximate characterization of the network steady-state. In the third category, \cite{Heterogeneous_Zhong17} extended the work in \cite{On_Zhong16} to the downlink cellular case under the same assumption of a fixed rate function and also derived approximations for the stability region. The authors in \cite{Spatiotemporal_Gharbieh17, Spatiotemporal_Gharbieh18,Self_Gharbieh19} focus on the characterization of the random access channel in an uplink cellular network, where different scheduling schemes were compared. Finally, \cite{Performance_Blaszczyszyn15} derives semi-analytic expressions for the stationary regime in a downlink cellular network.   To summarize, to the best of our knowledge, the stability of an uplink cellular system with an adaptive rate function is unknown. 

\subsection{Summary of Contributions}

In this paper, we study a single-cell uplink cellular system, where the users arrive at the base station's (BS) association area as a homogeneous Poisson point process (PPP) in time and space, each with a file with a random size to transmit to the BS. The service rate for each user depends on its current received SINR, i.e., $\log_2(1+{\rm SINR})$. Once the file is fully transmitted, the user leaves the network. First, we derive the exact necessary and sufficient condition for network stability. We show that this condition does not depend on the specific path loss function as long as the latter satisfies mild regularity and boundedness conditions. This condition is also oblivious of the fractional power control parameters and the power of the thermal noise.

Then we characterize the stationary regime when it exists. First, we propose a first-order approximation that has the interpretation of a mean-field limit. Precisely, the stationary distribution of the users' point process is assumed to follow an inhomogeneous PPP. We derive expressions for the network steady-state, and we show that this approximation is accurate in the low SNR regime, but loose otherwise. To this end, we propose a second-order approximation, which partially captures the correlations in the system. We demonstrate that this approximation accurately captures the true steady-state distribution of the network.

In the last part of this work, we discuss the metastability property of the system. In summary, we show that for a specific range of arrival rates that is higher than the critical threshold for stability, the system is unstable in the long run, but can be locally stable for a long time. Hence, in essence, it is similar to the instance of metastability observed in the Aloha case, and the SIS model \cite{liggett2013stochastic}. First, we discuss this property through the first-order approximation, since it has two solutions for its steady-state in this range of arrival rates. However, having multiple solutions for the mean-field model does not necessarily mean metastability. Hence, we analyze the mean first-passage time for the metastable range, which allows us to evaluate the mean time it takes the network to leave this locally stable regime and depart to the divergent one. We show that this time is very large compared to what happen in the unstable range. Furthermore, we show that this time grows linearly with the transmit power or the reciprocal of the thermal noise. Hence, it is a weaker form of metastability than the SIS model \cite{liggett2013stochastic}, for which this time grows exponentially with the model parameters.

 To the best of our knowledge, this is the first such case wireless network for  where metastability is not caused by the mobility of the users or the servers as in \cite{Metastability_Antunes06,Stochastic_Antunes08,Toward_Genin08,Metastability_Baccelli17}. We discuss why it is not observed in the ad hoc case but in the uplink cellular case. We also connect this observation to the metastability of slotted Aloha with exponential back-off \cite{Ultimate_Aldous87}. Note that, compared to wired networks and the works in \cite{Metastability_Antunes06,Stochastic_Antunes08,Toward_Genin08}, the mean-field limit approximation does not abstract the network (environment) as a single value that represents the network empirical average state. Instead, it abstracts the network state as a density function, which represents the spatial distribution of users in the network. Compared to \cite{Spatial_Sankararaman17}, which also uses a form of mean-field limit to analyze an ad hoc network, our analysis has to account for the user location in the cell, which changes the analytical approach and creates the phenomena described above. Also, in \cite{Spatial_Sankararaman17}, there are only two regions: stable and unstable. In our work, we have three regions: stable, metastable, and unstable.

The rest of the paper is organized as follows. In Section \ref{Sec:SysMod}, we present the system model. In Section \ref{Sec:MoA}, we discuss the reasons why this problem is challenging and we present our methodology of analysis. Section \ref{Sec:StabCond} is focused on deriving the necessary and sufficient condition of the network stability. Different approximations for the network steady-state regime are presented
in Sections \ref{Sec:FOA} and \ref{Sec:SOA}. Section \ref{Sec:Meta} is dedicated to metastability, which is studied from different perspectives and connected it to prior works on this topic. In Section \ref{Sec:Disc} we discuss further interpretations our  results and propose future research directions.%, before the conclusion in Section \ref{Sec:Conc}.

\section{System Model}\label{Sec:SysMod}
	We consider a single BS model, where the BS is located at the origin of the Euclidean space and has an association area defined by a compact set denoted by $\mathcal{D} \subset \mathbb{R}^2$.  Users arrive to $\mathcal{D}$ according to a homogeneous PPP in space and time with intensity $\lambda$  users per unit space and unit time. Hence, the number of users arriving in a region $\mathcal{A} \subset \mathcal{D}$ in a time period $T$ is a Poisson random variable with mean $\lambda T |\mathcal{A}|$, where $|\mathcal{A}|$ is used throughout this work to denote the area of region $\mathcal{A}$. Each user aims to transmit a file to the BS, and once the file is fully transmitted, the user leaves the network. Hence, our model represents  an uplink transmission in a single-cell cellular system or a multiple-access channel.
	
	The file sizes are assumed to be independent and identically distributed (i.i.d.) exponential random variables\footnote{This is a simplifying assumption since the file size typically follows a heavy-tailed distribution, as in the Internet case \cite{A_Crovella98}. However, it is necessary to maintain the Markovian property and the tractability of the model.} with mean $\frac{1}{\mu}$. The signal power attenuates with  distance according to a deterministic path loss function $L(\cdot)$. Small-scale fading is neglected. All active users transmit continuously on the same resource block (no scheduling) and interfere with each other. The transmit rate from a user to its BS at time $t$ is given by the rate function $R(x,\Phi_t)$, where $x$ is the location of the user and $\Phi_t$ is the set of the locations of active users at time $t$. We consider the following form for the rate function:
	\begin{align}\label{Eq:RateFunction}
	R(x,\Phi_t)=B \log_2 \left(1+\frac{P_xL(x)}{\sum\limits_{y\in \Phi_t\setminus\{x\}}P_yL(y)+\sigma^2}\right),
	\end{align}
	where $B$ is the bandwidth in Hz, $P_z$ is the transmit signal power of the user located at $z\in\mathcal{D}$, $L(z)$ is the path loss experienced by the signal, and $\sigma^2$ is the noise power. Hence, $P_x L(x)$ is the received power of the desired signal and  $\sum\limits_{y\in \Phi_t\setminus\{x\}}P_y L(y)$ is the interference power from all other users in the cell. So the rate function is the Shannon rate while treating interference as noise, where the BS is assumed to be able to decode the messages from all users perfectly since their rates are adapted based on their distance from the BS as well as the interference from other users. 
	
	We do not assume a specific shape for $\mathcal{D}$, but we consider a specific class of path loss functions called {\it Physically feasible path loss} models \cite[Definition 1]{A_AlAmmouri19}, where the path loss function has to be bounded and non-increasing: $L(x)\leq L(0)= L_{max} < \infty, \ \forall x\in \mathcal{D}$ to ensure that the received power is always finite and smaller than or equal to the transmit power.\footnote{The third property of the physically feasible path loss models mentioned in \cite[Definition 1]{A_AlAmmouri19} is always satisfied in our case since $\mathcal{D}$ is a compact set.} It was shown in \cite{A_AlAmmouri19} that, in addition to being physically necessary, this class of path loss functions includes a large variety of common path loss models that are used in the literature as well as in 3GPP standards. In addition, we assume that $L(x)\geq L_{min}>0, \ \forall x \in \mathcal{D}$, which is a reasonable assumption since if $L(x)=0, \ \forall x\in \mathcal{A}\subset \mathcal{D}$, then all users who arrive within $\mathcal{A}$ will not be served and  will accumulate, which leads to network instability. 
	
	Users are assumed to use fractional channel inversion power control \cite{Uplink_Xiao06,Fractional_Jindal08} and the transmitted signal of a user at location $x$ is $P_x=P L(x)^{l}$, where $l \in [0,1]$ is the channel inversion parameter. Hence, if $l=0$, all users transmit with the same fixed power, $P$, and if  $l=1$, then the users fully compensate for the path loss and the received power is constant for all users regardless of their locations. Note that due to the properties of the considered path loss model, namely $L(x)> 0$, the transmit power is ensured to be finite even with full channel inversion. 
	
	Our main focus in this work is on the low SINR regime, but we will comment on how to generalize our results to the general SINR case in Section \ref{Sec:Disc}. Note that in the low SINR regime, \eqref{Eq:RateFunction} reduces to:
	\begin{align}\label{Eq:RateFunction2}
	R(x,\Phi_t)=\frac{B}{\ln(2)}  \frac{ L(x)^{1-l}}{\sum\limits_{y\in \Phi_t\setminus\{x\}} L(y)^{1-l}+\bS},
	\end{align}
	where $\bS=\frac{\sigma^2}{P}$.
		
	\begin{definition}
		(Stability) The network is called stable if the number of active users converges weakly to a limit that does not depend on network initial condition.
	\end{definition} 
	
 Moreover, for stable networks, we are interested in characterizing their stationary regime (existence and uniqueness) and their ergodicity where the limiting empirical average fraction of time spent in a state is equal to the steady-state probability of being in that state.	In the next section, we discuss the main properties of our model along with the main mathematical tools used throughout this work. The notation is summarized in Table \ref{tb:Notation}.

	\begin{table}[t]
	\centering
	\caption{Notation.}
	\label{tb:Notation}
	{%
		\begin{tabular}{|l|l|}
			\hline
			\rowcolor[HTML]{EFEFEE} 
			Notation & Definition \\ \hline
			$\mathcal{B}(x,R)$ &  The disk centered at $x$ with radius $R$. \\ \hline
			$\mathcal{D}$ &  The association region of the BS. \\ \hline
			$\lambda$ &  The arrival rate in users per unit area and unit time. \\ \hline
			$\mu$ &  The reciprocal of the average file size. \\ \hline
			$P_z$ &  The transmit power, in Watts, of the user located at $z\in\mathcal{D}$.\\ \hline
			$\sigma^2$ & The average noise power in Watts. \\ \hline
			$B$ & The bandwidth in Hz. \\ \hline
			$l$ &  The channel inversion factor.\\ \hline
			$P$ & The transmit power scaling factor.\\ \hline
			$\bS$ &  $\frac{\sigma^2}{P}$.\\ \hline
			$\lambda_c$ &  The critical arrival rate. \\ \hline
			$\rho$ &  The loading factor, $\rho=\frac{\lambda}{\mu}$.\\ \hline
			$\Phi_t$ &  The set of the locations of active users at time $t$.\\ \hline
			$\Phi$ &  The stationary distribution of $\Phi_t$.\\ \hline
	\end{tabular}}
\end{table}

\section{Methodology of Analysis}\label{Sec:MoA}
Due to the Poisson arrivals and the exponential distribution of the file sizes, the system is Markovian: given the current state of the system, future states are independent of the previous states. In other words, the network can be modeled as a continuous-time Markov chain (CTMC), where the network state at time $t$ is captured by the locations of the users $\Phi_t$, which can be expressed as a counting measure $\Phi_t=\sum\limits_{i} \delta_{x_i}$, where $x_i$ is the location of the $i^{\rm th}$ user and $\delta(\cdot)$ is the Dirac measure. Hence, the system evolves with time as a spatial birth-death process\cite{Spatial_Preston75} defined on the state space of counting measures, and the invariant measure (the stationary distribution) is in the form of a random counting measure (a point process) if it exists.

Given that the users' positions at time $t$ are given by $\Phi_t$, the probability of a user arriving to the cell in the next tiny time period $\epsilon_t \ll 1$ is $\lambda |\mathcal{D}|\epsilon_t$, which is independent of the network state, and the probability that a user leaves within $\epsilon_t$ is given by 
\begin{align}\label{Eq:Proof1st}
\epsilon_t \mu \sum\limits_{x_i \in \Phi_t}R(x_i,\Phi_t) =\frac{\mu B\epsilon_t}{\ln(2)} \sum\limits_{x_i \in \Phi_t}\frac{L(x_i)^{1-l}}{\sum\limits_{x_j\in \Phi_t \setminus \{x_i \}}L(x_j)^{1-l}+\bS}.
\end{align}

From these equations, one can carry out the derivations of the exact transition probabilities and the Kolmogorov backward equations to prove (or disprove) the stability or the ergodicity of this CTMC and characterize its stationary regime \cite{Spatial_Preston75}. However, such an approach may not be tractable in our case due to the non-trivial form of the death rate.  From another perspective, one can think of our model as a multi-class single-server queuing model that employs a generalized processor sharing policy \cite{A_Parekh93}, where a user from the class $x$ gets a service rate as in \eqref{Eq:RateFunction2}. However, the number of different classes in our case is uncountably infinite (the continuum). Hence, in the following, we describe the main tools we used to tackle this problem.

First, note that if the system is not stable for a given arrival rate $\lambda_c \in \mathbb{R}_{+}$\footnote{We use $\mathbb{R}_{+}$ to denote the set of all non-negative numbers including zero.}, then it follows by monotonicity that the system is not stable for all $\lambda > \lambda_c$. Because a higher arrival rate leads to more users in the system, which increases the interference and reduces the transmission rate, which in turns increases the duration of stay of the users. Hence if the system is not stable for $\lambda_c$, it cannot be stable for all $\lambda > \lambda_c$. This can be rigorously proven by using a simple coupling argument. If we further assume that the system is stable for all $\lambda<\lambda_c$, then $\lambda_c$ is the critical arrival rate for which the system transitions from the stable regime to the unstable regime. Note that at this point, we do not assume that $\lambda_c$ is finite nor strictly positive; it can be $0$ hence the system is ultimately unstable regardless of the arrival rate, and it can be $\infty$ for which the system is stable for all finite arrival rates.

\begin{definition}\label{Def:Critical}
	(Critical arrival rate) The critical arrival rate, $\lambda_c$, is defined as the arrival rate which for all $\lambda > \lambda_c$ the network is not stable and for all  $\lambda <\lambda_c$, the network is stable regardless of the network initial condition.
\end{definition} 

In the next section, we follow a different approach than \cite{Spatial_Preston75} to prove the necessary and sufficient condition for the stability, where we derive the critical arrival rate $\lambda_c$ in a very simple form. Our approach is based on proposing other carefully designed CTMCs: one CTMC stochastically dominates our CTMC, and the other is stochastically dominated by our CTMC. Note that one of the advantages of this approach is the ability to transform our CTMC which takes values in the uncountable set of counting measures to other CTMCs that have a simpler structure and take their values in some countable set. This allows us to leverage the classical analysis of CTMCs defined on countable sets which have been widely studied in the literature.

After deriving the sufficient and necessary condition for the network stability and proving that the network admits a unique stationary regime if it is stable in Section \ref{Sec:StabCond}, we characterize its stationary regime in Sections \ref{Sec:FOA} and \ref{Sec:SOA}. Let $\Phi$ be the weak limit of $\Phi_t$ as $t\rightarrow \infty$ which represents the point process in the stationary regime assuming it exists, and let $\gamma(\cdot)$ be its intensity function (first order measure). In other words, the average number of users in a measurable set $\mathcal{A}\subset \mathcal{D}$ is $\int_{\mathcal{A}} \gamma(x)dx$ in the steady state. Our objective is to characterize $\Phi$ and its intensity function $\gamma(\cdot)$ as a function of the system parameters. In the following, we describe the approach we follow to achieve our goal.

First, note that if the system is stable, then the mean birth rate has to be equal to the mean death rate in the steady-state. In other words, the rate conservation principle \cite{Elements_Baccelli13} has to be satisfied in the stationary regime if the system is stable. The rate conservation principle in our case can be stated as:
\begin{align}
\frac{\lambda}{\mu}&=\gamma(x) \mathbb{E}\left[R(x,\Phi) | x\in \Phi \right], \ \ \ \ \ \forall x  \in \mathcal{D},\label{Eq:SC_2_0}
\end{align}
where the left hand side (LHS) represents the arrival rate (birth rate) at location $x$ in bps per unit area: $\lambda$ is in $\frac{\text{user}}{\text{m}^2 \text{sec}}$ and $\frac{1}{\mu}$ is in $\frac{\text{bits}}{\text{user}}$. Similarly, the right hand side (RHS) is the departure rate (death rate) which is also in bps per unit area: $\gamma (\cdot)$ is in $\frac{\text{user}}{\text{m}^2}$ and $R(x,\Phi)$ is in $\frac{\text{bits}}{\text{user} \ \text{sec}}$.  By substituting \eqref{Eq:RateFunction2} in \eqref{Eq:SC_2_0} we get 
\begin{align}
\rho&=\gamma(x) \frac{B}{\ln(2)} \mathbb{E}\left[ \frac{L(x)^{1-l}}{\sum\limits_{y\in \Phi\setminus\{x\}}L(y)^{1-l}+\bS} \Bigg| x\in \Phi \right], \ \ \ \ \ \forall x  \in \mathcal{D},\label{Eq:RateConsPri}
\end{align}
where $\rho=\frac{\lambda}{\mu}$ is in bps per unit area.

Note that the expectation in \eqref{Eq:RateConsPri} is w.r.t. the point process $\Phi$ which has an intensity function $\gamma(\cdot)$. Hence, $\gamma(\cdot)$ has two opposite effects in \eqref{Eq:RateConsPri}: higher $\gamma(\cdot)$ increases the term outside the expectation, but it also increases the denominator inside the expectation, which represents the network interference power. %In other words, if we ignore the dependency of the function inside the expectation on $\gamma(\cdot)$ (fixed per-user rate), then higher $\gamma(\cdot)$ increases the cumulative departure rate (death rate) and if we ignore the term outside the expectation, then higher $\gamma(\cdot)$ decreases the term inside  the expectation (the per-user throughput) due to the interference. 
Hence, \eqref{Eq:RateConsPri} captures the inter-dependency between the queue status of the users and their service rates. This inter-dependency makes the system hard to analyze exactly, especially given that the point process type $\Phi$ is not known, and obtaining $\gamma(\cdot)$ alone may not be sufficient because higher order moment measures are also needed to evaluate the expectation in \eqref{Eq:RateConsPri} as we will show in the next sections. Hence, we propose different approximations and heuristics that have the flavour on mean-field limits to analyze this network in Sections  \ref{Sec:FOA} and \ref{Sec:SOA}, and we rely on simulations to show the accuracy of these approximations.

\section{Stability Conditions}\label{Sec:StabCond}
	In this section, our objective is to provide the necessary and sufficient condition for the network stability, derive the critical arrival rate, and prove that the network admits a unique stationary regime when it is stable. Overall, the results are summarized in the following theorem.
	
	\begin{theorem}\label{Th:ThmStab}
		The cutoff arrival rate for the CTMC $\Phi_t$, as defined in Definition \ref{Def:Critical},  is 
		\begin{equation}
		    \lambda_c=\frac{B \mu}{\ln(2) |\mathcal{D}|},
		\end{equation}
		 users per unit area and unit time. More precisely, the CTMC is ergodic (stable) with a unique stationary distribution for all $\lambda<\lambda_c$, and transient (unstable) for all $\lambda>\lambda_c$.  
	\end{theorem}
	
    Note that the stability condition is independent of the specific path loss function, the channel inversion parameter $l$, and the noise power. The latter is expected since when the network operates close to the critical threshold, a large number of active users are expected to be present all the time. This leads to the domination of interference over noise in the denominator of the rate function \eqref{Eq:RateFunction2}. The independence from the path loss function and the channel inversion factor will be clear in the next section. In summary, the network adapts to the path loss and the channel inversion through the density function of the active users; higher path loss leads to higher density, and smaller channel inversion factor also leads to higher density. However, the network does not transition from the stable to the unstable regimes by just changing the path loss or the channel inversion parameter.
	
	To prove Theorem \ref{Th:ThmStab}, we start by dividing the region $\mathcal{D}$ into $N_{\epsilon}$ disjoint connected sets $A^{(\epsilon)}_{j}, j\in \{ 1, 2, \cdots, N_{\epsilon}\}$ with equal areas $ \epsilon=\frac{|\mathcal{D}|}{N_{\epsilon}}$. Such a tessellation is possible since the region $\mathcal{D}$ is compact. Furthermore, define the following:
\begin{align}
\bar{L}_i^{(\epsilon)}=\sup_{x \in A^{(\epsilon)}_{i}} L(x)^{1-l},\notag\\
\barbelow{L}_i^{(\epsilon)}=\inf_{x \in A^{(\epsilon)}_{i}} L(x)^{1-l}\notag.
\end{align}

Since $L(\cdot)$ is continuous, non-increasing, and bounded from below and above by $L_{\rm min}>0$ and $L_{\rm max}<\infty$, respectively, we have the following
\begin{align}
\lim\limits_{\epsilon \rightarrow 0}\barbelow{L}_i^{(\epsilon)}&=\lim\limits_{\epsilon \rightarrow 0} \bar{L}_i^{(\epsilon)},\\
\lim\limits_{\epsilon \rightarrow 0} \frac{\bar{L}_i^{(\epsilon)}}{\barbelow{L}_i^{(\epsilon)}}&=\lim\limits_{\epsilon \rightarrow 0} \frac{\barbelow{L}_i^{(\epsilon)}}{\bar{L}_i^{(\epsilon)}}=1.\label{Eq:Thm1_lim1}
\end{align}

Define the CTMC $\barbelow{\Phi}$ which counts the number of nodes in each region $A_{i}^{(\epsilon)}$. Hence, at each time instant, $\barbelow{\Phi}$ is a $1 \times N_{\epsilon}$ vector $[k_i]_{i=1}^{N_{\epsilon}}$, where $k_i\in \mathbb{N}$ is the number of nodes in the region $A_{i}^{(\epsilon)}$ and the CTMC takes values in the countable set $\{\mathbb{N}\}^{N_{\epsilon}}$. The arrival rate for each region $A_{i}^{(\epsilon)}$ is $\lambda \epsilon$, which means that the total arrival rate over all regions is the same as the arrival rate to the original process $\Phi$. Define the service rate for a node located in the $i^{\rm th}$ region given that $\barbelow{\Phi}=[k_j]_{j=1}^{N_{\epsilon}}$ and $k_i\geq1$ as:
\begin{align}\label{Eq:ServRate}
\frac{B\mu}{\ln (2)} \frac{\barbelow{L}^{(\epsilon)}_i}{\sum\limits_{j=1}^{N_{\epsilon}}k_j\bar{L}^{(\epsilon)}_j+\bS}.
\end{align}

Moreover, define the CTMC $\bar{\Phi}$ similarly to $\barbelow{\Phi}$, except that the service rate for a user located in the $i^{\rm th}$ region given that $\barbelow{\Phi}=[k_j]_{j=1}^{N_{\epsilon}}$ and $k_i\geq1$ is:
\begin{align}
\frac{B\mu}{\ln (2)} \frac{\bar{L}^{(\epsilon)}_i}{(k_i-1)\barbelow{L}^{(\epsilon)}_i+\sum\limits_{\substack{j=1\\j\neq i}}^{N_{\epsilon}}k_j\barbelow{L}^{(\epsilon)}_j+\bS}.
\end{align}

\begin{lemma} \label{Lm:CTMCs}
	Based on the definitions of the CTMCs $\Phi$, $\barbelow{\Phi}$, and $\bar{\Phi}$, we have the following:
	\begin{enumerate}
		\item The CTMC $\Phi$ is $\phi$-irreducible.
		\item The CTMC $\barbelow{\Phi}$ stochastically dominates the CTMC $\Phi$. 
		\item The CTMC $\bar{\Phi}$ is stochastically dominated by the CTMC $\Phi$. 
	\end{enumerate}
	
	\begin{proof}
		The irreducibility \cite{Markov_Meyn12} of $\Phi$  can be shown be picking a measure that has a unit mass at the empty state (no active users) and zero elsewhere. The stochastic dominance \cite{Markov_Meyn12} proofs are based on simple coupling arguments. For the full proof, refer to Appendix \ref{App:CTMCs}.
	\end{proof}
\end{lemma}

Hence, $\barbelow{\Phi}$ stochastically dominates $\Phi$, which implies that a sufficient condition for the stability of $\barbelow{\Phi}$ is also a sufficient condition for the stability of ${\Phi}$. 
In next theorem, we derive a sufficient condition for the stability of $\barbelow{\Phi}$.

\begin{theorem}\label{Th:Nec}
	For all $\lambda$ such that $\lambda<\frac{B \mu}{\ln(2) |\mathcal{D}|}$, the CTMC $\barbelow{\Phi}$  is ergodic with a unique stationary regime.
	\begin{proof}
		The proof relies on the Foster-Lyapunov Theorem \cite[theorem 5.1.1]{Markov_Pierre13}, where we show that for an appropriate Lyapunov function, the drift of $\barbelow{\Phi}$  is negative outside a compact set and finite inside it. For the full proof, refer to Appendix \ref{App:Thm2}.
	\end{proof}
\end{theorem}

Since $\barbelow{\Phi}$ stochastically dominates the CTMC $\Phi$ and $\Phi$ is $\phi$-irreducible, it follows that $\Phi$ is also ergodic with a unique stationary regime if $\lambda<\frac{B \mu}{\ln(2) |\mathcal{D}|}$.  For the uniqueness of the stationary regime, we need in addition the $\phi$-irreducibility of $\Phi$.

To complete the proof of Theorem \ref{Th:ThmStab}, we need  to show that $\Phi$ is unstable for $\lambda>\frac{B \mu}{\ln(2) |\mathcal{D}|}$. For this, we consider the stability of $\bar{\Phi}$ in the next theorem.

\begin{theorem}\label{Th:Suf}
	The CTMC $\bar{\Phi}$ is transient (unstable) for all $\lambda>\frac{B \mu}{\ln(2) |\mathcal{D}|}$.
	\begin{proof}
		Refer to Appendix \ref{App:Thm3}.
	\end{proof}
\end{theorem}

Since  $\bar{\Phi}$ is stochastically dominated by $\Phi$, a  necessary condition for the stability of $\bar{\Phi}$ is also necessary for the stability of ${\Phi}$. Hence, we can conclude by the last theorem and Lemma \ref{Lm:CTMCs} that  $\Phi$ is transient (unstable) for all $\lambda>\frac{B \mu}{\ln(2) |\mathcal{D}|}$, which completes the proof of Theorem \ref{Th:ThmStab} and concludes this section.

\section{Stationary Regime: First Order Approximation}\label{Sec:FOA}

After proving that the network has a unique stationary regime, we shift our focus to characterizing this stationary regime.  Hence, unless otherwise stated, the network is assumed to be operating in the stable region, i.e., the arrival rate is less than the critical threshold given in Theorem \ref{Th:ThmStab}. %Note that since the CTMC is defined on the state space of counting measures, its invariant (stationary) distribution we are interested in, takes the form of a random counting measure or a point process \cite{Stochastic_Baccelli10} which represents the probability of the presence of a user at any location $x\in \mathcal{D}$ in the stationary regime.
The main tool we use in this section is the rate-conservation principle which is given in \eqref{Eq:RateConsPri}. Our objective is to characterize  $\Phi$ through its moment measures. However, the expectation in \eqref{Eq:RateConsPri} is w.r.t. $\Phi$, which we are trying to analyze and on which we do not know anything, except that it exists. To overcome this issue,  our approach is to assume certain structural properties   for $\Phi$ that help us simplify the form in  \eqref{Eq:RateConsPri} and then find the desired intensity function. We will also discuss the intuition behind such approximations and compare their accuracy by comparison to the results we get from our simulator. 

In this section, we assume that $\Phi$ follows an inhomogeneous PPP with intensity function $\gamma_p(\cdot)$. Note that the intensity function (first-moment measure) fully characterizes a PPP due to its independence property; higher order moment measures can be found by the intensity function, hence the name {\it First Order Approximation}. However, our model is expected to have some correlations as we will study later in this section.

 One can also get this approximation using a mean-field limit. As we mentioned, in the mean-field limit, each queue evolves independently from the current state of the environment; the status of other queues, and it only observes the empirical average state of the environment.  Hence, one can abstract the environment by the interference term in \eqref{Eq:RateConsPri}. In that case, each user, regardless of its location, observes the average interference, i.e., $\mathbb{E} [I]$. Another finer abstraction of the network leverages  the location of the users. In this case, the user observes the first measure of the stationary distribution of the users in the cell, i.e., $\gamma(\cdot)$, and its service rate does not depend on the current users in the network. Since this applies to each user in the network, all users evolve independently from each other, which leads to the independence property of the PPP. Overall, the mean-field limit in this case also leads to a PPP with intensity function $\gamma_p(\cdot)$. This is the intuition behind the mean-field approximation, for a more formal definition, refer to Appendix \ref{App:MeanF}.  

Despite the lack of correlations in the PPP assumption, this approximation reveals the metastability property of this model;  this is one of the reason behind discussing this approximation in detail. 
\subsection{Main Results}
 The stationary regime under the first order approximation is fully described by the following theorem.% by solving for $G(\gamma_p)$, we get the  equilibrium points. Then by substituting in \eqref{Eq:SC_5}, we get the intensity function $\gamma_p(\cdot)$. 

\begin{theorem}\label{Th:FOA}
	Under the first order approximation, if the system is stable for a certain set of parameters, then the intensity function is given by 
	\begin{align}
	\gamma_p(x)&=\frac{Z^{*}}{L(x)^{1-l}},\label{Eq:SC_IntFun_1}
	\end{align}
	where $Z^{*}$ is a solution for the following fixed point equation
	\begin{align}
	\frac{\rho \ln(2)}{B} = Z \int\limits_{0}^{\infty} e^{-t \bS} \exp \left(-Z \int\limits_{\mathcal{D}} \left( 1- e^{-t L(y)^{1-l}}\right) L(y)^{l-1} \mathrm{d}y \right) \mathrm{d}t. \label{Eq:SC_8_1}
	\end{align}
	\begin{proof}
Refer to Appendix \ref{App:FOA}.
	\end{proof}
\end{theorem}

Hence, under this approximation, the stationary regime is captured by a single-variable fixed point equation, which can be easily evaluated numerically.  A special case of interest for Theorem \ref{Th:FOA} is the case of full power control, i.e., $l=1$, which we present in the next corollary.

\begin{corollary}
	For the special case of full channel inversion, $l=1$, the intensity function  $\gamma_p^{*}$ is constant. This constant is a solution of the following fixed-point equation
	\begin{align}
	&\frac{\rho \ln(2)}{B}= \gamma e^{-\gamma |\mathcal{D}| } \int\limits_{0}^{\infty}  \exp \left( -t \bS +\gamma |\mathcal{D}|  e^{-t }  \right) \mathrm{d}t. \label{Eq:SC_8_2}
	\end{align}
% 	\begin{align}
% 	    &\frac{\rho \ln(2)}{B}= \gamma e^{-\gamma |\mathcal{D}| } \int\limits_{0}^{1} y^{\bS-1} \exp \left(\gamma |\mathcal{D}|  y  \right) dy.\\
% 	    &\frac{\rho \ln(2)}{B}= \gamma e^{-\gamma |\mathcal{D}|} \left( -\gamma |\mathcal{D}| \right)^{\bS } \left( \Gamma(\bS ) -\Gamma(\bS -\gamma |\mathcal{D}|)\right).
% 	\end{align}
% 	\begin{proof}
% 	$y=\exp(-t), t=-\log(y),\mathrm{d}t=-1/y\mathrm{d}t$
% 	\end{proof}
\end{corollary}
	Hence, for this case, the stationary regime simplifies to a homogeneous PPP with intensity $\gamma$. This result is expected since the service rate for each user is independent of its locations, i.e., all users are stochastically identical. 
	
\subsection{Analysis}
Now we analyze the expressions we derived for the network stationary regime under this approximation. Note that the stationary regime in \eqref{Eq:SC_8_1} and \eqref{Eq:SC_8_2} takes the form of a fixed point equation. Ideally, we would like these equations to have a unique solution  when the network is stable and to not have any finite solutions when the network is unstable. To this end, we start with the simple case of $l=1$, where the intensity function is flat and the average number of users in the cell in the stationary regime simplifies to $\bar{N}=\gamma |\mathcal{D}|$. Define $C:=\frac{\rho \ln(2) |\mathcal{D}|}{B}$ and $f(\bar{N}):=\bar{N} e^{-\bar{N} } \int\limits_{0}^{\infty}  \exp \left( -t \bS +\bar{N}  e^{-t }  \right) \mathrm{d}t$. Hence, the fixed point equation in \eqref{Eq:SC_8_2} can be written as $C=f(\bar{N})$. The properties  of this fixed point equation is given in the next corollary.
		
\begin{corollary}\label{Cor:SolutionsFOA}
		For a fixed $C$, the number of solutions only depends on $\bS$. Namely,
		    \begin{itemize}
		       \item For any strictly positive value of  $\bS$, $f(0)=0$ and $\lim\limits_{\bar{N}\rightarrow \infty}f(\bar{N})=1$.
		        \item If $\bS \geq 1$, then the fixed point equation has a unique solution if $C \in [0,1]$ and no solution if $C> 1$. For the special case of $\bS=1$, this solution is given by $C=1-\exp(-\bar{N})$. 
		        \item If $\bS <1$, then we have three cases:
		        \begin{itemize}
		            \item If $C\in[0,1)$, then the equation has a unique solution.
		            \item If $C\in(1,C_1]$, where $ \max(\frac{e^{-1}}{\bS},1)\leq C_1 \leq 0.5^{\bS-1}\left(1+\frac{1}{\bS}\right)$, then the equation has two solutions.
		            \item If $C\in(C_1,\infty)$, then the equation has no solutions.
		        \end{itemize}
		    \end{itemize}
		   \begin{proof}
		       The sketch of the proof is as follows: for the case of $\bS \geq 1$, we prove that $f(\bar{N})$ is strictly increasing for any finite and positive $\bar{N}$, and since $\lim\limits_{\bar{N}\rightarrow \infty}f(\bar{N})=1$, this proves the statement in the corollary. For the case of $\bS < 1$, we prove that $f(\bar{N})$ is strictly increasing if $0\leq\bar{N}\leq1$ and strictly decreasing if $\frac{\bS+1}{\bS-1}\leq\bar{N}<\infty$, which means that for a certain range of  $\bar{N}$, $f(\bar{N})$ is larger than one since $\lim\limits_{\bar{N}\rightarrow \infty}f(\bar{N})=1$, and it also means that $f(\bar{N})$ has at least one local maximum value. Then we proceed and prove that $f(\bar{N})$ has a single local maximum  that is less than $\frac{1}{2}^{\bS-1}\left(1+\frac{1}{\bS}\right)$, which concludes the proof. For the details, refer to Appendix~\ref{App:proofSolutionsFOA}.
		   \end{proof}
		\end{corollary}

Hence, the fixed-point equation has a unique solution when the network is stable, which is desired. It also does not have any finite solution when the network is unstable and is operating in the low SNR regime. However, we might have two solutions in the high SNR regime, even if the network is unstable. Before discussing the meaning of this, we study the general fractional power control case to see whether such a phenomenon exists in the general case as well. To this end, we consider a numerical example, given that the fixed-point equation in Theorem \ref{Th:FOA} does not have a simple form. Precisely, we assume that the association area $\mathcal{D}$ is a disk centered at the origin with radius $R$. For the path loss function, we take $L(r)=(1+r)^{-\eta}$, where $\eta$ is the path loss exponent. Under these assumptions, the network has the following parameters: $\lambda,\mu, l, R,\eta,\bS$ and $B$. We fix $\mu=\frac{1}{100}$ bits$^{-1}$, $B=1$ MHz, $\bS=-50$ dBm, $R=100$ meters and we vary $\lambda$, $l$, and $\eta$. Note that since the intensity function is inversely proportional to the path loss as shown in \eqref{Eq:SC_IntFun_1}, the network steady-state is fully captured by the variable $Z^{*}$ which is a solution to the fixed point equation in \eqref{Eq:SC_8_1}. Equivalently, one can look at the average number of active users $\bar{N}$ in the stationary regime, since it has a clear physical  meaning. The average number of users in the stationary regime is given by

 \begin{align}\label{Eq:SC_avgN_1}
 \bar{N}=\int\limits_{\mathcal{D}} \gamma_p(x) \mathrm{d}x=Z^{*} \int\limits_{\mathcal{D}} L(x)^{l-1} \mathrm{d}x.
 \end{align}
 
     \begin{figure*}[t]
		\centering
		\begin{subfigure}{\textwidth/2}
			\centerline{\includegraphics[width=  3.2in]{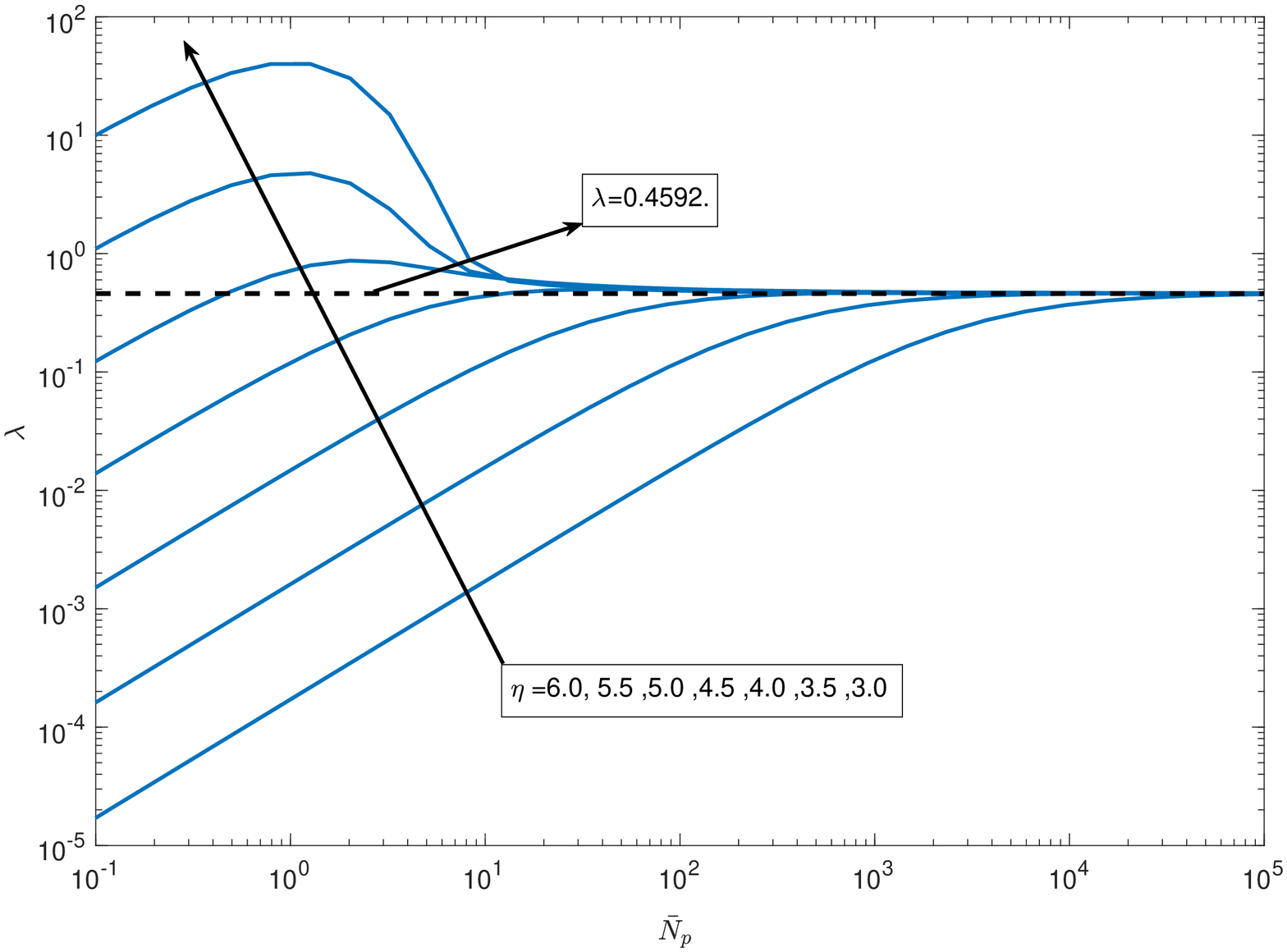}}
			\caption{\, Fixed $l=0$ and different $\eta$.}
			\label{fig:LvsN}
		\end{subfigure}%
		\begin{subfigure}{\textwidth/2}
			\centering
			\centerline{\includegraphics[width=  3.2in]{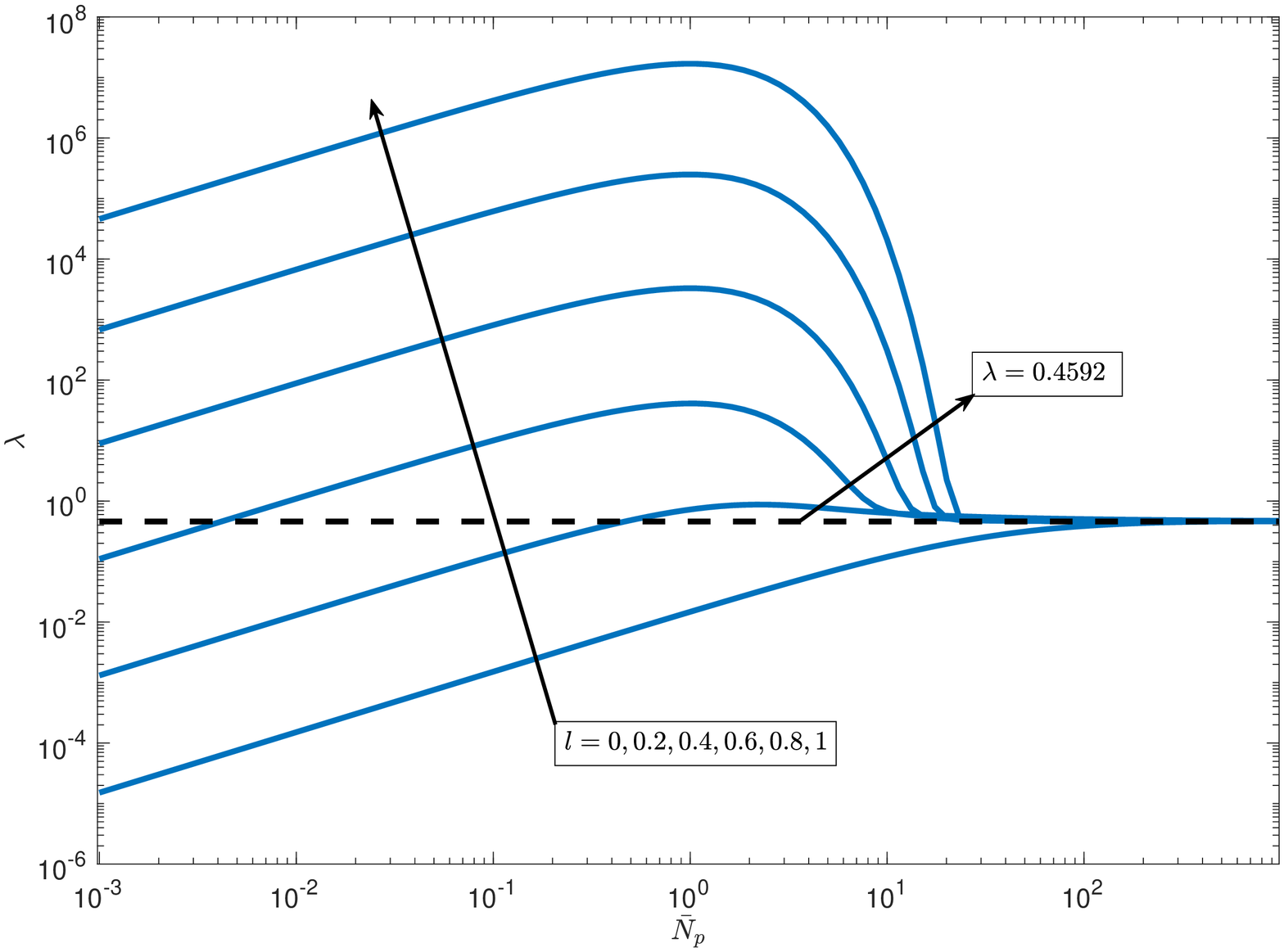}}
			\caption{\, Fixed $\eta=5$ and different $l$.}
				\label{fig:LvsN_PC}
		\end{subfigure}
		\caption{\,  The arrival rate ($\lambda$) vs. the average number of users ($\bar{N}$) for different path loss exponents and channel inversion parameters under the first order approximation. The dashed line is the critical threshold from Theorem \ref{Th:ThmStab}.}
		\label{fig:stabilityFOA}
	\end{figure*}
    
 We start by showing how $\bar{N}$ varies with the arrival rate, the channel inversion parameter,  and the path loss exponent. For each value of  $\eta$, $l$ and $\bar{N}$, we find the corresponding $Z^{*}$ from \eqref{Eq:SC_avgN_1} and then by plugging $Z^{*}$ in \eqref{Eq:SC_8_1}, we find $\lambda$. The results are shown in Fig. \ref{fig:stabilityFOA}. First, note that when the average number of users tends to infinity, the arrival rate tends to the critical threshold, which is consistent with Theorem \ref{Th:ThmStab} and similar to the conclusion of Corollary \ref{Cor:SolutionsFOA}.

Moreover, the fixed-point equation has a single solution as long as the arrival rate is less than the critical threshold given in Theorem \ref{Th:ThmStab}. However, similar to  Corollary \ref{Cor:SolutionsFOA}, the figures show an interesting behavior for small path loss exponents or full channel inversion. They show that for some $\lambda\geq\lambda_c$, where $\lambda_c$ is the critical threshold mentioned in Theorem 1, the fixed point equation in \eqref{Eq:SC_8_1} has two solutions. Note that we proved that the network is unstable in this region and this result is based on an approximation, and hence it might seem like an overkill to analyze this behavior. However, given that this approximation has a mean-field limit interpretation, having multiple solutions could mean that the network is \textit{metastable} \cite{Stochastic_Antunes08}. To maintain the flow of the paper, we delay discussing metastability to the last section and we move next to assess the accuracy of the first order approximation in the stable case.
 
\subsection{Accuracy of The First Order Approximation}
First, we describe the simulation setup. Time is discretized into tiny intervals of $\epsilon_t$, where $\epsilon_t$ is set such that the average number of users who arrive to the whole cell within $\epsilon_t$ is $1/100$ users. Hence, $\epsilon_t$ is different for different arrival rates. For each time slot, the number of users who arrive to the network is a realization of a Poisson random variable with mean $\lambda |\mathcal{D}| \epsilon_t$, where these users are uniformly scattered within $\mathcal{D}$ each equipped  with a file size that is exponentially distributed with mean $\frac{1}{\mu}$ bits. At the end of each time slot, the transmit rate for each user is calculated based on its SINR, and the transmitted bits are subtracted from its file size. Once the user finishes transmitting its file, it leaves the network. The total number of time steps is set to be $10^7$, and the results are averaged over at least $10$ different realizations of the network (using different seeds for the random variables).

%This can be observed in Fig. \ref{fig:FOA2}, where the network behaves as if it is stable for a long time. However, due to the infrequent large fluctuations of the arrivals, it escaped this equilibrium point and the number of users started growing rapidly. 

\begin{figure}[t]
	\centerline{\includegraphics[width=  4.5in]{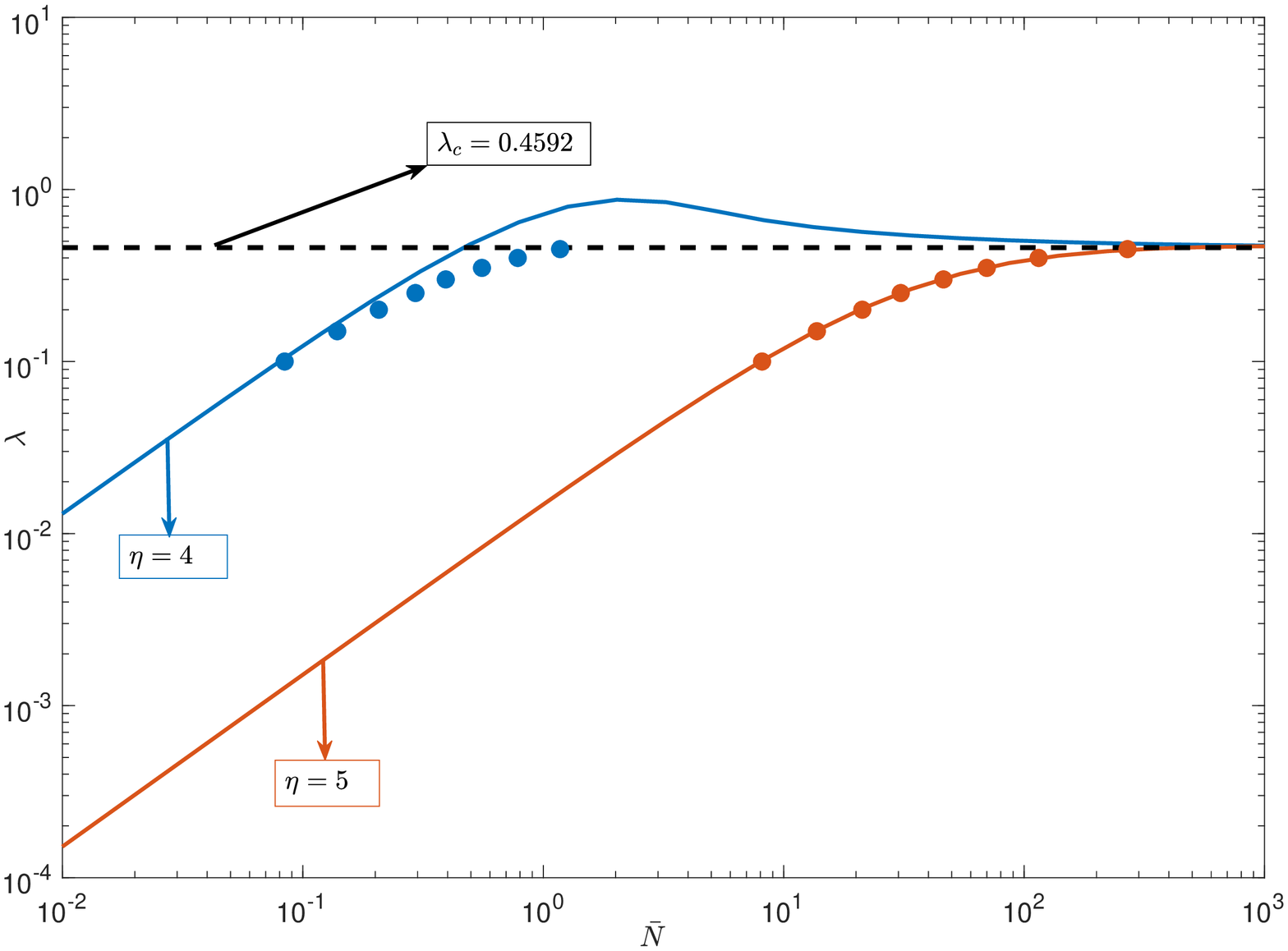}}
	\caption{\,  The arrival rate ($\lambda$) vs. the average number of users ($\bar{N}$). The solid lines are the results based on the first order approximation and the filled circles are the results by simulations. The curves are for $\eta=4$ and $\eta=5$.}
	\label{fig:stabilityCir}
	\end{figure}

To capture the network evolution with time for a fixed $\lambda$, $l$, and $\eta$, we focus on the number of active users $\bar{N}_t$ at each time step and then we average over all time steps to get the average number of users in the cell $\bar{N}$. The results are shown in Fig. \ref{fig:stabilityCir}, where the solid line is the first order approximation and the filled circles are the results from the simulator. Note that since increasing $l$ has the same effect as decreasing $\eta$, as shown in Fig. \ref{fig:FOA}, we fix $l$ to $0$ hereafter and vary $\eta$.

First, note that although we assumed an inhomogeneous PPP, the intensity function only captures the effect of the distance between the user and its BS through the path loss function since we showed that the intensity function is inversely proportional to the path loss function after inversion. However, it does not capture any correlation between the users' locations; the number of users in disjoint sets in $\mathcal{D}$ are independent in the first order approximation.  As Fig. \ref{fig:stabilityCir} shows, the first order approximation is accurate for the cases of high path loss exponent and/or the small arrival rate. This is because the interference in these cases is not dominant, hence neglecting the correlation has a negligible effect.
    
    However, the figure shows that the first order approximation is loose for small path loss exponents and high arrival rates. This is because interference is dominant in these cases and the bottleneck is no longer the desired signal power. The correlation between the users' locations can be explained through the following simple example. Assume that many users are observed near the BS; this means that these users are suffering from high interference and are not exiting fast enough. Hence, it is expected that cell-edge users also have trouble exiting the network because they are additionally suffering from the low desired signal power due to the long distance between them and the BS. This means that observing many users near the BS implies that one should observe many far users also. This correlation between the users' locations cannot be captured through the first order approximation since the number of nearby users is independent of the number of far users in this approximation. 
    
    Overall, the first-order approximation is simple and has a clear intuition, but as we have shown, it is loose when the network interference power is dominant. In the next section, we propose another approximation that partially captures the correlations in the interference term.

\section{Stationary Regime: Second Order Approximation}\label{Sec:SOA}
 
To capture the correlations between the locations of the users, we have to characterize the higher order moment measures of $\Phi$. Let $\gamma^{(2)} (x_{k},x_{v})$ denote the second moment measure of the users' point process \cite{Stochastic_Baccelli10}, which represents the mean of the inner product of the number of users at $x_{k}$ and the number of users at $x_{v}$:
\begin{align}\label{Eq:SecOrdMea}
    \mathbb{E}\left[  N(\mathcal{A}_1) N(\mathcal{A}_2)\right]=\int_{\mathcal{A}_1} \int_{\mathcal{A}_2}\gamma^{(2)} (x,y)\mathrm{d}x\mathrm{d}y, 
\end{align}
where $\mathcal{A}_1,\mathcal{A}_2 \subset\mathcal{D}$. One can further define the higher order measures in the same manner, i.e., $\gamma^{(3)} (\cdot,\cdot,\cdot)$,  $\gamma^{(4)} (\cdot,\cdot,\cdot,\cdot)$, etc. Note that in the first order approximation, we assumed that $\Phi$ is fully characterized by its first moment measure, i.e., $\gamma^{(1)} (\cdot)$, and hence, a single fixed point equation was enough to find it. However, to fully capture the true system performance, one would need a fixed point equation per moment measure. To this end, we assume in this section that $\Phi$ is fully defined by its first two moment measures, i.e., $\gamma^{(2)} (\cdot,\cdot)$ and $\gamma^{(1)} (\cdot)$, and hence the name {\it second} order approximation. In this case, it is assumed that higher order measures can be factorized into the first two moment measures similar to \cite{Can_Baccelli13}, as we discuss in detail in Appendix \ref{App:SOA}. After using this assumption along with a few others, which we discuss in  Appendix \ref{App:SOA} to maintain the flow of the paper, we get the following theorem.
 
\begin{theorem}\label{Th:SOA}
    Under the second order approximation, the steady-state distribution of the point process $\Phi$ is characterized by the following coupled fixed point equations for all $x,y \in \mathcal{D}$.
    \begin{align}
&\!\!\!\!\!\!\!\!\!\frac{\rho \ln(2)}{B} = \gamma^{(1)} (x)L(x)^{1-l} \int\limits_{0}^{\infty} e^{-\bS t} \exp \left(-\int\limits_{\mathcal{D}} \left( 1- e^{-t L(y)^{1-l}}\right) \frac{\gamma^{(2)} (y,x)}{\gamma^{(1)}(x)} \mathrm{d}y \right) \mathrm{d}t, \label{Eq:SOH_9_1}\\
&\!\!\!\!\!\!\!\!\!\frac{\rho \ln(2)}{B}    \left(\gamma^{(1)}(x)+\gamma^{(1)}(y) \right)=\notag\\
& \!\!\!\!\!\!\!\!\!\gamma^{(2)}(x,y) \left(  \frac{L(x)^{1-l}}{ \int\limits_{\mathcal{D}}\frac{\gamma^{(3)}(x,y,u)}{\gamma^{(2)}(x,y)} L(u)^{1-l}\mathrm{d}u+L(y)^{1-l}+\bS}+ \frac{L(y)^{1-l}}{\int\limits_{\mathcal{D}}\frac{\gamma^{(3)}(x,y,u)}{\gamma^{(2)}(x,y)} L(u)^{1-l}\mathrm{d}u+L(x)^{1-l}+\bS}\right),\label{Eq:SOH_9_2}
\end{align}
where $\gamma^{(3)}(x,y,u)$ is factorized as any convex combination of the following: $\gamma^{(2)}(x,y)\gamma^{(1)}(u)$, $\gamma^{(2)}(x,u)\gamma^{(1)}(y)$, $\gamma^{(2)}(y,u)\gamma^{(1)}(x)$, and $\gamma^{(2)}(y,u)\gamma^{(2)}(x,u)\frac{1}{\gamma^{(1)}(u)}$.
\begin{proof}
Refer to Appendix \ref{App:SOA}.
\end{proof}
\end{theorem}

\begin{figure}[t]
	\centerline{\includegraphics[width=  4.5in]{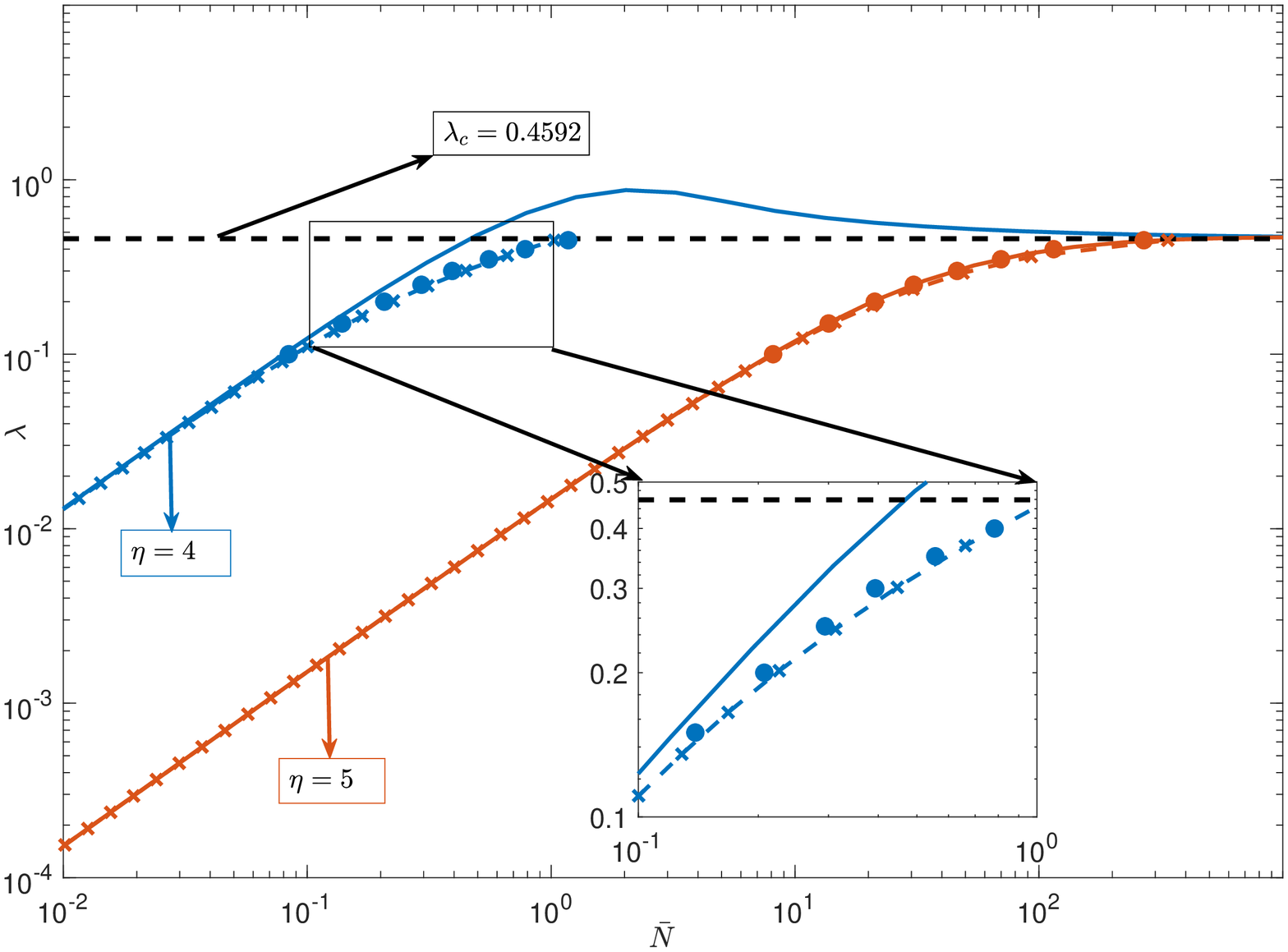}}
	\caption{\,  The arrival rate ($\lambda$) vs. the average number of users ($\bar{N}$). The solid lines are the results based on the first order approximation, the dashed lines marked with 'x' are the results based on the second order approximation and the filled circles are the results by simulations. The curves are for $\eta=4$ and $\eta=5$.}
	\label{fig:stabilityCir3}
\end{figure}

Hence, the system performance is captured by two coupled fixed point equations. To evaluate the accuracy of this approximation, we use the same parameter values as in the previous section, and we solve these two fixed point equations as follows. Both $\gamma^{(1)}(\cdot)$ and $\gamma^{(2)}(\cdot,\cdot)$ are initialized assuming the first order approximation. Then, $\gamma^{(2)}(\cdot,\cdot)$  is found iteratively using \eqref{Eq:SOH_9_2}. Then $\gamma^{(1)}(\cdot)$ is found iteratively using equation \eqref{Eq:SOH_9_1}.  The results are shown in Fig. \ref{fig:stabilityCir3}, showing that this approximation matches well with the simulation results for high and low path loss and arrival rate. 

\begin{figure}[t]
	\centerline{\includegraphics[width=  4.5in]{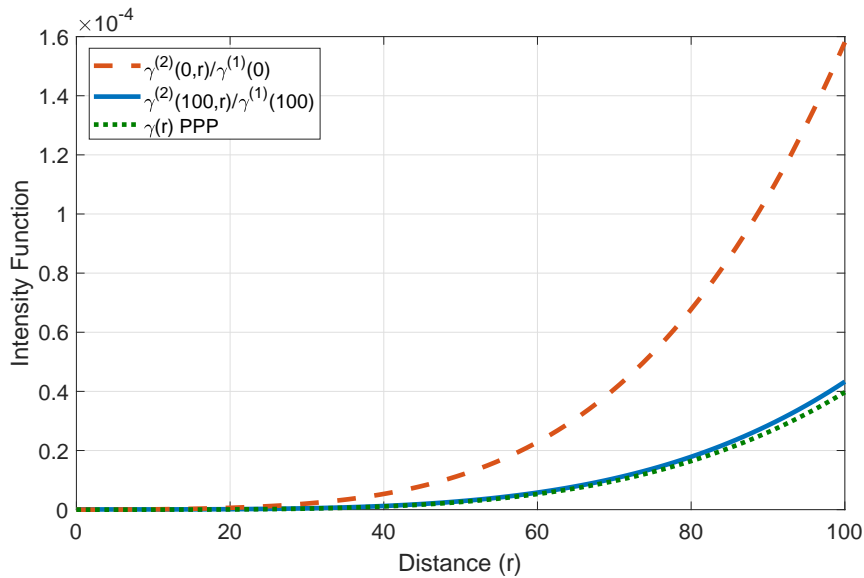}}
	\caption{\, The intensity function under different approximations for $\lambda=0.425$ and $\eta=4$.}
	\label{fig:diffInt}
\end{figure}

Before wrapping up this section, we check whether the intensity function we get from the second order approximation matches with our intuition discussed earlier. We already mentioned that observing nearby users increases the probability of the presence of cell edge users, which we believe to be the reason why the first order approximation is loose in regions where the interference is dominant. In Fig. \ref{fig:diffInt}, we plot the intensity function of the user point process seen by a user located at the origin in a red dashed curve, the intensity function of the user point process seen by a cell-edge user in a solid blue curve, and the intensity function that we get from the first order approximation, which assumes that different users observe the same intensity function of the other users PP, in a dotted green curve. As the figure shows, the first order approximation underestimates the intensity function compared to the second order approximation regardless of the location of the observer.

Moreover, given that the observer is at the origin, it sees a higher intensity function compared to the cell-edge user, which agrees with our intuition. Note that this result roughly means that the probability for a cell-center to observe a cell-edge user is four times the probability for a cell-edge to observe another cell-edge user, which is significant.  Hence, thanks to the second order approximation, we have a tight approximation for the steady-state distribution of our point process $\Phi$, and we have already proved its stability in the previous section. Overall, we now have  a full heuristic characterization of the point process $\Phi$.

\section{Metastability}\label{Sec:Meta}
So far, we proved that the network has a unique stationary regime if the arrival rate is less than the critical threshold given in Theorem \ref{Th:ThmStab} and we have presented two different approximations to characterize this stationary regime. In this section, we focus on the case when the network is unstable, $\lambda>\lambda_c$. As we mentioned, one might question the reason for studying this region, given that the network is unstable in this case, which means that, eventually, we will have an infinite accumulation of users in network. The answer to this questions is metastability.

As we mentioned in the introduction, in a specific case of metastability, the system has locally stable points, and it can stay in their neighbourhoods for a very long time before departing to its absorbing state as in \cite{Metastability_Baccelli17} and the SIS model in \cite{liggett2013stochastic}. It appears that our model exhibits a similar property with an absorbing state at $\infty$, and hence in is unstable in the long-run, despite being locally stable for a long time. To understand this property, we look at it from three different approaches; the first-order approximation, discrete event simulation, and finally  first passage times analysis. 

\subsection{Analysis}
\subsubsection{First-order approximation} 
\begin{figure}[t]
		\centerline{\includegraphics[width=  4in]{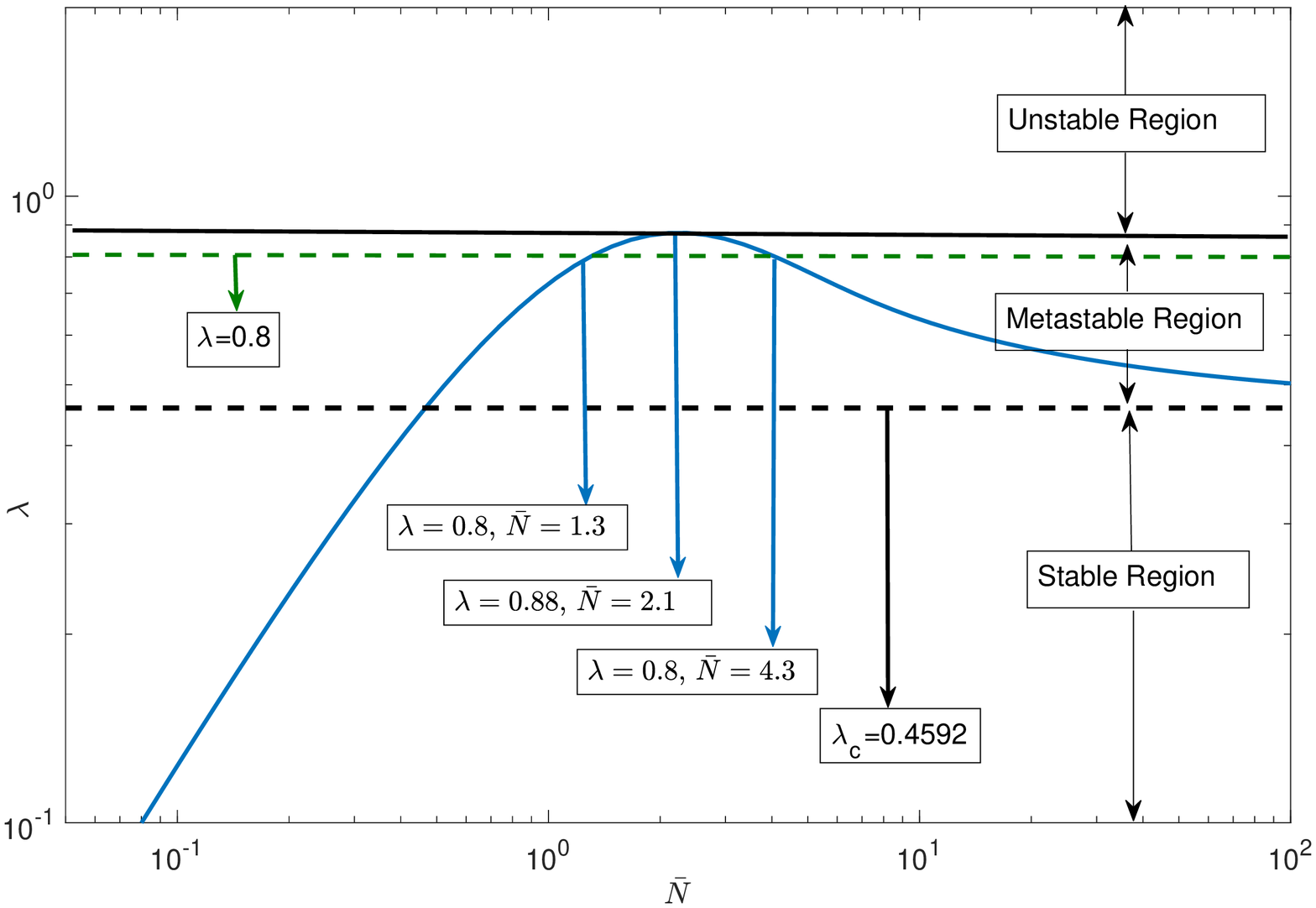}}
		\caption{\, $\lambda$ vs $\bar{N}$ under the first order approximation, where $\eta=4$ and $l=0$.}
		\label{fig:FOA}
\end{figure}

Looking back at the first order approximation, we proved that the fixed point equation has two solutions for a range of $\lambda>\lambda_c$. This could mean that the network is metastable is this range of arrival rates, although it is not necessary. To explain this, we consider the example of $\eta=4$ and $\lambda=0.8$ shown in Fig. \ref{fig:FOA}. In this case, in the network has two  equilibrium points, i.e., solutions to the fixed point equation, the first at $\bar{N}=1.3$ and the second at $\bar{N}=4.3$. At these points, the mean birth rate is equal to the mean death rate. Assume that the network is operating at the first equilibrium point. Following the curve, an increase (decrease) in the arrival rate around this point, increases (decreases) the average number of users in the network. However, this is not the case around the second equilibrium point: a decrease (increase) in the arrival rate, increases (decreases) the average number of users in the network. Hence, the system does not react properly around this point, which leads to instability. 

Note that the system might operate around the first equilibrium point for a long time and act as if it were stable, until, due to the randomness in the arrivals, it passes the peak at $\bar{N}=2.1$ due to the arrival of many users at short period of time. Afterwards, users start accumulating, and the system descends to instability. Based on this, the system, at least under this mean-field model, has a locally stable point, which it stays in its neighbourhood, given that we do not have huge fluctuations in the arrivals.\footnote{Although we focused on the case where these infrequent large fluctuations are caused by the arrival of many users, there are other reasons as well. For example, the arrival of a few users, but each with a huge file to transmit, or the arrival of many cell-edge users.}

\subsubsection{Simulations} 
\begin{figure}[t]
	\centerline{\includegraphics[width=  4.5in]{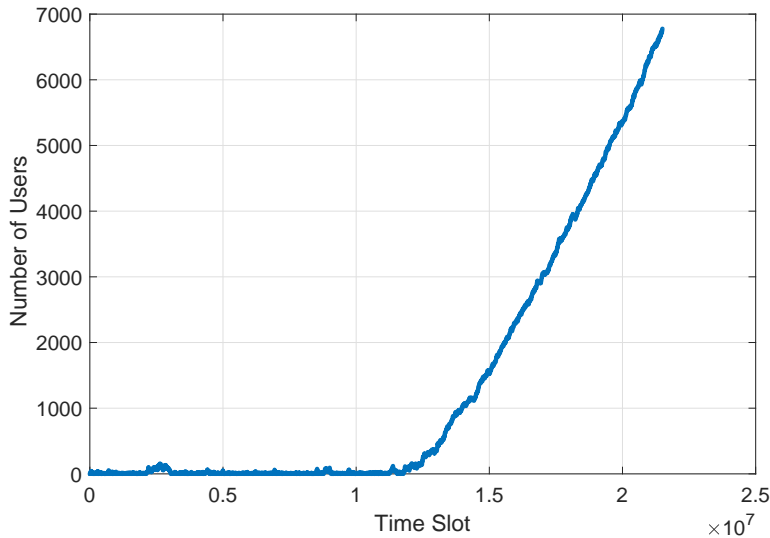}}
	\caption{\,  The evolution of the number of users with time for the metastable case.}
	\label{fig:Unstable}
	\end{figure}

To observe this phenomenon in practice, we use a discrete event simulation. An example from our simulator (which we described in detail earlier) is shown in Fig. \ref{fig:Unstable}, where we plot the evolution of the number of users with time in a  case where $\lambda$ is in the metastable window. Note that the network behaves as if it was stable for a long time, and although there is a jump in the number of users around the time step $2.5\times 10^6$, it came back to its meta-stable regime. However, around time step $1.25\times 10^7$, the number of users starts growing linearly, due to one of these large infrequent fluctuations in the arrival rate, and the network starts drifting  to $\infty$.
	
\subsubsection{First Passage Times} To have a better quantitative understanding of the metastability, we focus on the case of full channel inversion ($l=1$). In this case, all users are treated equally and the spatial aspect of the network is eliminated since the service rate of a user is independent of its location and the location of other users in the network. Hence, the network evolves with time as a traditional one dimensional birth-death Markov chain, where the states represent the total number of users in the network. Moreover, the arrival rate to any state is $\lambda |\mathcal{D}|$, regardless of the number of users in that state and the departure rate of the $n^{\rm th}$ state is

	\begin{align}
	    \frac{B\mu}{\ln(2)} \frac{n}{(n-1) \mathbbm{1} (n>1)+\bS},
	\end{align}
	where $\bS=\frac{\sigma^2}{P}$. From Theorem \ref{Th:ThmStab}, the system is not stable for all arrival rates higher than  $\frac{B \mu}{\ln(2) |\mathcal{D}|}$. However, as we will show next, the system can temporarily handle an arrival rate higher than $\frac{B \mu}{\ln(2) |\mathcal{D}|}$. Precisely, as long as the network operates within a compact set of states, the departure rate is higher than the arrival rate, which leads to a negative drift, and the network acts as if it was stable. This consistent to what we discussed about the system having a locally stable point. To show this, we set the arrival rate to $\frac{B \mu}{\ln(2) |\mathcal{D}|} (1+\epsilon)$, where $\epsilon>0$. Then to have a negative drift, the following needs to be satisfied
	\begin{align}
	      \frac{n}{(n-1)+\bS} \geq (1+\epsilon),
\end{align}
	      {or equivalently,}
	   \begin{align}
	      n \leq \frac{1+\epsilon}{\epsilon}(1-\bS),\label{eq:Meta11}
	\end{align}
	where $\bS\leq\frac{1}{1+\epsilon}<1$. Hence, as long as the number of users in the network is less than $\frac{1+\epsilon}{\epsilon(1-\bS)}$, we have a negative drift and the network acts as if it was stable. However, due to the Poisson arrivals, a large number of users can (will eventually) arrive within a tiny period of time, which will push the network state outside the compact set in \eqref{eq:Meta11}. Then the number of users starts to grow towards infinity due to the positive drift, which leads to instability of the network.
	
	Recall that $\bS=\frac{\sigma^2}{P}$, and hence, the size the compact set scales with the transmit power $P$, and hence, the mean time the network operates within this locally stable compact set is expected to scale with the transmit power. Also, note that this compact set only exists if $\bS<1$. This is consistent with the first-order approximation, since we showed that the fixed point equation only can have two solution if $\bS<1$. Hence, these two arguments are well connected so far and both indicate the existence of a locally stable regime if $\bS<1$.
	%Note that the mean-field analysis gives an indication to the metastability in this case as shown in Fig.~\ref{fig:LvsN_PC} since we have two equilibrium points. A similar argument holds for the general case, where we have no or partial channel power inversion by dividing the region $\mathcal{D}$ into disjoint connected sets as in Section \ref{Sec:StabCond} and then work out the negative drift condition in terms of the number of users within each region.} 

    The final piece in the puzzle is the time needed to leave this compact set. Hence, We complement the previous analysis by studying the expected time required for the network to blow up. Let $\tau_{m,n}$, where $m<n$, be the time required to have $n$ points in the system starting from the state of $m$ points, i.e., $\tau_{m,n}=\inf\limits_{j}\{ X_j=n|X_0=m\}$, where $X_j$ is the number of points in the system at time step $j$. Since the spatial aspects are neglected in the case of $l=1$, one can follow \cite{A_Keilson65} and express the first moment of $\tau$ as
\begin{align} \mathbb{E}[\tau_{n,n+1}]&=\frac{1+\frac{n}{n-1+\bS}\mathbb{E}[\tau_{n-1,n}]}{\lambda|\mathcal{D}|},\\
	\mathbb{E}[\tau_{0,n}]&=\sum\limits_{j=0}^{n-1}\mathbb{E}[\tau_{j,j+1}],
		\end{align}
	where $n\geq1$ and $\mathbb{E}[\tau_{0,1}]=\frac{1}{\lambda |\mathcal{D}|}$ and ${\text var}(\tau_{0,1})=\frac{1}{(\lambda |\mathcal{D}|)^2}$. More explicit expressions are given in the following lemma.

		\begin{lemma}
		    For an arrival rate of $\lambda |\mathcal{D}|=1+\epsilon$, where $\epsilon>0$, the first moment  of the first passage time can be expressed as follows.
		    \begin{align}
		   \mathbb{E}[\tau_{n,n+1}]&=\frac{1}{1+\epsilon}+\sum\limits_{i=1}^{n}\frac{1}{(1+\epsilon)^{i+1}}\frac{\Gamma(n+1)\Gamma(n+\bS-i)}{\Gamma(n+\bS)\Gamma(n+1-i)}, \label{eq:PTONE}\\
		   \mathbb{E}[\tau_{0,n}]&=\frac{n}{1+\epsilon}+\sum\limits_{k=0}^{n-1}\sum\limits_{i=1}^{k}\frac{1}{(1+\epsilon)^{i+1}}\frac{\Gamma(k+1)\Gamma(k+\bS-i)}{\Gamma(k+\bS)\Gamma(k+1-i)}.\label{eq:PTCum}
		\end{align}
		 For the special case of $\bS=1$, these quantities simplify to the following 
		 \begin{align}
		   \mathbb{E}[\tau_{n,n+1}]&=\frac{1}{\epsilon}-\frac{1}{\epsilon (1+\epsilon)^{n+1}}, \label{eq:PTONE1}\\
		   \mathbb{E}[\tau_{0,n}]&=\frac{\epsilon n-1}{\epsilon^2}+\frac{1}{\epsilon^2(1+\epsilon)^{n}}.\label{eq:PTCum1}
		\end{align}
		Moreover, all these quantities are decreasing with $\bS$.
		\begin{proof}
		The expressions can be easily verified by induction. The decreasing property follows by rewriting  \eqref{eq:PTONE} as
		\begin{align}
		  \mathbb{E}[\tau_{n,n+1}]&=\frac{1}{1+\epsilon}+\sum\limits_{i=1}^{n} \frac{1}{(1+\epsilon)^{i+1}}\frac{(n)_i}{(n-1+\bS)_i},
		\end{align}
		where $(n)_i=\prod\limits_{j=0}^{i-1}(n-j)$ is the falling factorial.
		\end{proof}
		\end{lemma}

\begin{figure*}[t]
		\centering
		\begin{subfigure}{\textwidth/2}
			\centerline{\includegraphics[width=  3.2in]{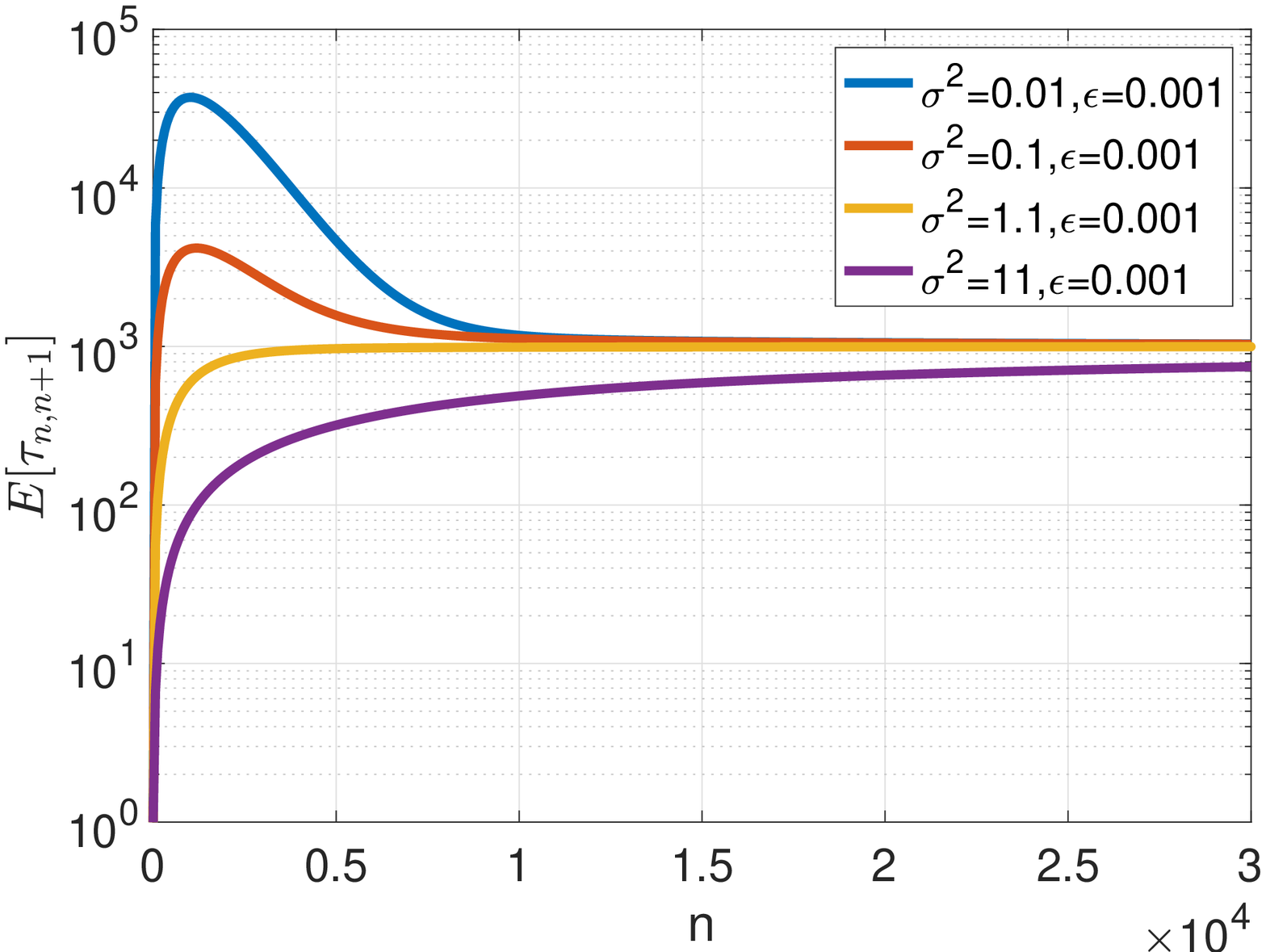}}
			\caption{\, The mean one-step passage time vs $n$.}
			\label{fig:OnePT}
		\end{subfigure}%
		\begin{subfigure}{\textwidth/2}
			\centering
			\centerline{\includegraphics[width=  3.2in]{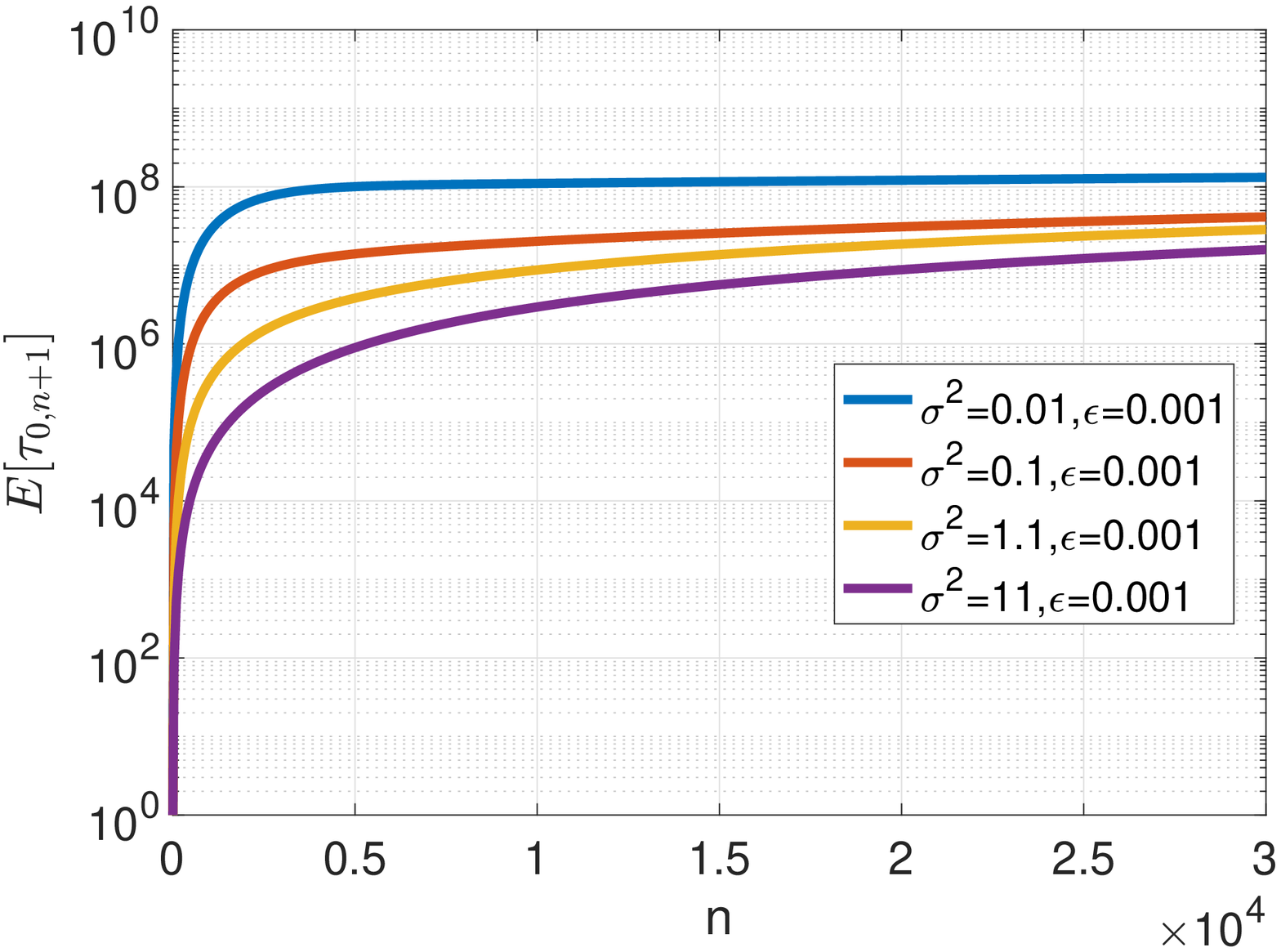}}
			\caption{\, The total passage time vs $n$.}
				\label{fig:FPT}
		\end{subfigure}
		\caption{\,  The mean first passage time.}
	\end{figure*}

Hence, the time needed to observe a large number of users in the network is larger in the case of $\bS<1$ compared to the case of $\bS\geq1$. However, the previous lemma does not state how much larger. To this end, we plot the one-step and the total passage time in Fig. \ref{fig:OnePT} and Fig. \ref{fig:FPT}, respectively, for these two cases. There are three main observations from these figures. First of all, regardless of $\bS$, the mean one-step passage time saturates to the constant $\frac{1}{\epsilon}$. Hence, asymptotically, the average time needed to observe a large number of users in the system scales linearly. This is consistent with Theorem \ref{Th:ThmStab}, since we know that the network is unstable in long-run, hence, the newly arriving users will get stuck in the network due to the low service rate.   

The second observation is that $\mathbb{E}[\tau_{0,n}]$ is significantly larger for the case of $\bS<1$. For example, the time need to observe $3 \times 10^4$ users in the network has a mean of $1.1 \times 10^8$ and $2.9\times 10^6$ for the cases of $\bS=0.01$ and $\bS=11$, respectively, which justifies what we observed through our simulations. The third observation is that $\mathbb{E}[\tau_{0,n}]$ grows at a rate faster than linear for intermediate values of $n$ in the case of $\bS<1$. In fact, using basic curve fitting, we found that it grows as $n^2$ initially before slowing down to $n$ for large values of $n$. This holds for any value of $\bS<1$, but the scaling factor depends on the chosen value of $\bS$, the smaller the value, the larger the scaling factor. However, the range over which it scales as $n^2$ seems to be independent of $n$ and only based on $\epsilon$. 

\begin{figure}[t]
		\centerline{\includegraphics[width=  4in]{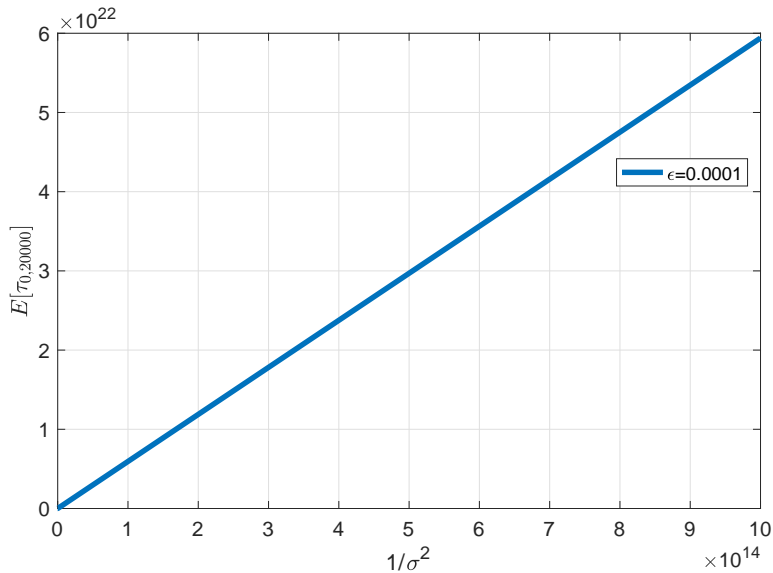}}
		\caption{\, $\mathbb{E}[\tau_{0,n}]$ vs $\frac{1}{\bS}$ for $n=20000$.}
		\label{fig:LastFig}
\end{figure}

Hence, so far we have shown that there is a locally stable regime if $\bS<1$, and the time it takes the system to depart from it and diverge is potentially large. To wrap up the analysis, we study how the mean first passage time scales with $\bS$, which we plot in Fig. \ref{fig:LastFig}. As the figure show, $\mathbb{E}[\tau_{0,n}]$, for large $n$, scales {\it linearly} with $\frac{1}{\bS}$. Note that in systems where metastability is observed as in the SIS model \cite{liggett2013stochastic}, or in \cite{Metastability_Antunes06,Stochastic_Antunes08,Toward_Genin08} and the Ising model in statistical physics \cite{liggett2013stochastic}, $\mathbb{E}[\tau_{0,n}]$ scales exponentially with some network parameters, e.g., the network size. In our case, it scales linearly when the network is metastable. Hence, our model exhibits a weaker form of metastability with all qualitative properties as in the classical models, but for growth rate of the time to instability which is linear rather than exponential.

\subsection{Connection to prior works}	
	Note that a similar property was observed in discrete-time slotted Aloha \cite{Data_Bertsekas92}. However, in that case, the network is  unstable regardless of the arrival rate \cite{Ultimate_Aldous87}. This is related to our model as follows: both our model and slotted Aloha describe a network where multiple nodes transmit to a common receiver. Hence, they are very similar from this perspective. However, our interference model is way softer, and more accurate, than the collision model for Aloha. In other words, our model tolerates more interference. Hence, we do not always have ultimate instability, but we still have metastability in the case of severe interference. 
	
	{Another interesting line of work where metastability was observed is \cite{Metastability_Antunes06,Stochastic_Antunes08,Toward_Genin08}. The basic underlying model for these works consists of a finite number of queues (BSs) with finite capacity with multiple classes of users arriving uniformly to these queues. Users migrate from one queue to another according to some exponential time clock if the chosen queue has the capacity to accommodate them, and otherwise, they leave the network. It was shown in \cite{Metastability_Antunes06,Stochastic_Antunes08,Toward_Genin08} that for some network parameters, the network also has two globally stable points, where the network state remains at one of these points for a long time, then due to a rare event, the state switches to the other equilibrium point and stays there also for a long time. Hence, in this case, the metastability is different from the one we observe, where the network becomes unstable. Metastability in \cite{Metastability_Antunes06,Stochastic_Antunes08,Toward_Genin08} is believed to be due to three ingredients: $(i)$ mobility, $(ii)$ multiple cells, and $(iii)$ multiple classes of users. Interestingly, in our case, we have only a single cell without mobility. Also, we observe metastability in the case of full power control, $l=1$, where there is no discrimination against cell-edge users and we have only one class of users. Hence, the cause of metastability is different from that in \cite{Metastability_Antunes06,Stochastic_Antunes08,Toward_Genin08} and it is closer in this sense to the metastability of Aloha, since there is also a single class of users there. Next, we move to study the accuracy of the first order approximation with the aid of our simulator.

\section{Discussion and Future Work}\label{Sec:Disc}
\subsection{Discussion}
First, note that we have focused so far on the low SINR regime, where the rate function is given by \eqref{Eq:RateFunction2}. However, the results we found can be easily extended to the general rate function given by \eqref{Eq:RateFunction}. For example, the next theorem shows that the stability region given in Theorem \ref{Th:ThmStab} does not change by switching to the general rate function.

\begin{theorem}\label{Thm:GeneralRa}
    Under the general rate function given in \eqref{Eq:RateFunction}, the cutoff arrival rate for the CTMC $\Phi_t$ is
    \begin{equation} \label{Eq:Thm6_1}
		    \lambda_c=\frac{B \mu}{\ln(2) |\mathcal{D}|},
		\end{equation}
		 users per unit area and unit time. More precisely, the CTMC is ergodic (stable) with a unique stationary distribution for all $\lambda<\lambda_c$, and transient (unstable) for all $\lambda>\lambda_c$.
		 \begin{proof}
		 Refer to Appendix \ref{App:Thm:GeneralRa}.
		 \end{proof}
\end{theorem}

Note that operating close to the critical arrival rate means that the users are having a hard time flushing out from the network since a small increase in the average arrival rate can lead to instability of the network. This implies that the users are experiencing very low SINR and the rate function in \eqref{Eq:RateFunction} approaches the one in \eqref{Eq:RateFunction2} for low SINR, i.e.,  $\ln(1+{\rm SINR}) \approx {\rm SINR}$. Hence, it is not surprising that the critical arrival rate under the general rate in \eqref{Eq:RateFunction} is the same as in Theorem \ref{Th:ThmStab}. The first and second order approximations under the general rate function can also be derived following similar approaches as in Sections \ref{Sec:FOA} and \ref{Sec:SOA}. However, these derivations do not reveal more insights than what we have already discussed.

The next point we discuss in this section is the critical arrival rate. Note that this threshold, given in Theorem \ref{Th:ThmStab}, is independent of the chosen path loss. However, the intensity function of the users is inversely proportional to the path loss. Hence, one can think of this threshold as a constraint on the cumulative rate that flushes out of each tiny piece of $\mathcal{D}$. In other words, pick $\mathcal{A}\subset \mathcal{D}$, then increasing the path loss experienced by the users within $\mathcal{A}$ decreases the service rate for these users, but at the same time increases the number of users in this region, such that the product of the number of users within $\mathcal{A}$ and their service rate is always kept constant. Hence, changing the path loss does not lead to instability of the network, since the network adapts to it by increasing the intensity function and reducing the per-user service rate.

\subsection{Future work}
Our next focus is on extending this model to the multi-cell case, where in addition to intra-cell interference we have inter-cell interference. This additional interference correlates the state of the different cells in the network, in addition to the correlation between the users within the same cell we observe in the single cell case. Hence, this extension requires a substantial amount of work, and we postpone it for future work.
%Replica method and metastability \cite{Topics_Sznitman91,Stochastic_Antunes08,A_Bordenave07,Queuing_Baccelli13,A_LeBris12,The_Aristoff15}

%General SINR
% \section{Conclusion}\label{Sec:Conc}
%  {In this paper, we have analyzed the performance of a single-cell uplink cellular system with traffic dynamics. Specifically, we derived the necessary and sufficient condition for the network stability, which is oblivious of the specific path loss as long as it satisfies some natural boundedness and monotonicity conditions. Then we show through mean-field analysis and verify through our simulations that there is a regime where the network has two equilibrium points in the mean-field limit. If the network is operating around the first one, it acts as if it is stable. But, if due to infrequent large fluctuations of the arrivals, the network is  pushed to operate beyond the second point, it becomes unstable. This property is referred to as metastability and was not observed in this wireless context. We also provide a heuristic characterization of the network steady-state when it stable using mean-field analysis and higher moment measures factorization, and we show that it is tight using simulations.}

\appendices
\section{Proof of Lemma \ref{Lm:CTMCs}}\label{App:CTMCs}
\subsection{Irreducibility of $\Phi$}
The irreducibility  of $\Phi$ can be shown by picking the measure $\phi$ to be the Dirac measure at the empty state (the state with no users), then by applying Theorem 4.0.1 in \cite{Markov_Meyn12} to the embedded chain of $\Phi$, we can deduce that it is $\phi$-irreducible. More specifically, we can get from any state that has $M$ nodes to the empty state in $M$ steps with a non-zero probability since the death rate of $\Phi$ is non-zero and the birth rate is finite. Namely, let the locations of the users be given by the set $\{x_i\}_{i=1}^{M}$, then the probability to be in the empty state after $M$ steps is lower bounded by

\begin{align}
\prod\limits_{i=1}^{M}{ \frac{q_i}{q_i+p}}>0,
\end{align}
where 
\begin{align}
q_i&=\frac{\mu B}{\ln(2)}  \frac{L(x_i)^{1-l}}{\sum\limits_{j=i+1}^{M} L(x_j)^{1-l}+\bS}>0,\\
p&=\lambda |\mathcal{D}|.
\end{align} 

 Hence, with non-zero probability, the return time to the empty state from any other state is finite which implies that the embedded chain is $\phi$-irreducible by \cite[Theorem 4.0.1]{Markov_Meyn12}. Finally, we can conclude that  $\Phi$ is also  $\phi$-irreducible according to \cite[Definition 7.2.1]{Stochastic_Gallager13}.

\subsection{Stochastic Dominance}
In the following, we prove the second point, that $\barbelow{\Phi}$ stochastically dominates the CTMC $\Phi$. The third point follows using the same approach. First, note that:
\begin{align}
&\bar{L}^{(\epsilon)}_i \geq L(x)^{1-l} \geq \barbelow{L}^{(\epsilon)}_i, \ \forall x \in A_{i}^{(\epsilon)},\label{Eq:proof1stobs}\\
&(k_i-1)\bar{L}^{(\epsilon)}_i+\sum\limits_{\substack{j=1\\j\neq i}}^{N_{\epsilon}}k_j\bar{L}^{(\epsilon)}_j< \sum\limits_{j=1}^{N_{\epsilon}}k_j\bar{L}^{(\epsilon)}_j.\label{Eq:proof2ndobs}
\end{align}

Hence, for a given set of nodes in $\mathcal{D}$, the death rate of each node under $\barbelow{\Phi}$ is smaller than the death rate under $\Phi$. This is clear from comparing \eqref{Eq:ServRate} with \eqref{Eq:Proof1st} taking into account  \eqref{Eq:proof1stobs} and \eqref{Eq:proof2ndobs}.  Moreover, if $\Phi \subseteq \barbelow{\Phi}$, then for each $x \in \Phi$, the death rate of $x$ under $\Phi$ is also smaller than the death rate of the same point under $\barbelow{\Phi}$ due to  \eqref{Eq:proof1stobs}, \eqref{Eq:proof2ndobs} and the possible increase in the denominator in \eqref{Eq:ServRate} because of the possible extra points in  $\barbelow{\Phi}$ (extra interference). Hence, if we start the processes $\Phi$ and $\barbelow{\Phi}$ in the same initial condition and we couple their arrivals (both see the same arrivals with the same files sizes), then the number of nodes in $\Phi$ is less than the number of nodes in $\barbelow{\Phi}$ throughout the whole trajectory of time. This can be explained by the following argument.

Start both of the Markov processes $\Phi$ and $\barbelow{\Phi}$ with the same initial condition, the same set of points in $\mathcal{D}$ with their file sizes, and then couple their arrivals such that the position of the new node and its file size is the same for both Markov processes. Hence, both $\Phi$ and $\barbelow{\Phi}$ has the same set of points until the first event occur $E_{1}$, which could be due to the following reasons:
\begin{enumerate}
	\item $A_{x}$: an arrival at position $x \in \mathcal{D}$.
	\item $D_{x}$: a departure of the point at $x \in \mathcal{D}$ in $\Phi$.
	\item $\barbelow{D}_{x}$: a departure of the point at $x \in \mathcal{D}$ in $\barbelow{\Phi}$.
\end{enumerate}

It is clear that if $E_{1}$ is of the type $A_{x}$ or $D_{x}$, then the number of nodes in $\Phi$ is still less than or equal to the number of nodes in $\barbelow{\Phi}$, so the ordering is maintained and $\Phi \subseteq \barbelow{\Phi}$. But, if the event $\barbelow{D}_{x}$ occurs, then the ordering is broken. However, given that the death rate in $\Phi$ of a point at $x$ is higher than the death rate in $\barbelow{\Phi}$ of the same point as we explained earlier, then a point cannot die in $\barbelow{\Phi}$ before it dies in $\Phi$. Hence, only the events $A_{x}$ or $D_{x}$ can occur and the ordering $\Phi \subseteq \barbelow{\Phi}$ is maintained in either case. Now consider the second event $E_{2}$ which can be $A_{y}$, $D_{y}$, or $\barbelow{D}_{y}$. As in the previous case, $A_{y}$ or $D_{y}$ maintain the ordering $\Phi \subseteq \barbelow{\Phi}$. But $\barbelow{D}_{y}$ can only occur if $E_{1}=D_{x}$ and $x=y$ because we showed that after the first event, the ordering $\Phi \subseteq \barbelow{\Phi}$ is maintained and we showed previously that if $\Phi \subseteq \barbelow{\Phi}$, then the death rate of each point in $\Phi$ is higher than $\barbelow{\Phi}$.

 Hence, a death can only occur for a point that already died in $\Phi$ and still alive in $\barbelow{\Phi}$. Which means that after the second event, the ordering $\Phi \subseteq \barbelow{\Phi}$ is still maintained. At this point, it is straightforward to show by induction that the ordering $\Phi_t \subseteq \barbelow{\Phi}_t$ will be maintained throughout the whole trajectory of time. Hence, the CTMC $\barbelow{\Phi}$ stochastically dominates the CTMC $\Phi$. 
\section{Proof of Theorem 2}\label{App:Thm2}

For simplicity, we study the embedded chain of $\barbelow{\Phi}$ denoted by $\barbelow{\Phi}^{(d)}$, where the superscript $(d)$ is used to denote that it is defined over discrete time. Define the following:
\begin{align}
p_i&=\lambda \epsilon,\\
p&=\sum\limits_{i=1}^{N_{\epsilon}}p_i=\lambda |\mathcal{D}|,\label{Eq:arravalRate}\\
q_i&=	\frac{B\mu}{\ln (2)} \frac{k_i\barbelow{L}^{(\epsilon)}_i}{\sum\limits_{j=1}^{N_{\epsilon}}k_j\bar{L}^{(\epsilon)}_j+\bS},\\
q&=\sum\limits_{i=1}^{N_{\epsilon}}q_i=	\frac{B\mu}{\ln (2)} \frac{\sum\limits_{i=1}^{N_{\epsilon}}k_i\barbelow{L}^{(\epsilon)}_i}{\sum\limits_{j=1}^{N_{\epsilon}}k_j\bar{L}^{(\epsilon)}_j+\bS}.
\end{align}

The transition probabilities of $\barbelow{\Phi}^{(d)}=[k_i]_{i=1}^{N_{\epsilon}}$ are non-zero only to states that have one more  unit (node) in one of the coordinates of the vector $[k_i]_{i=1}^{N_{\epsilon}}$, or one less unit (node). More specifically, the transition probabilities for the element $k_i$ are $k_i \rightarrow k_i+1 \ w.p. \ \frac{p_i}{p+q}$ and  $k_i \rightarrow k_i-1 \ w.p. \ \frac{q_i}{p+q}$. Based on these probabilities, it is clear that $\barbelow{\Phi}^{(d)}$ is irreducible and aperiodic: from any configuration that has $M \in \mathbb{N}$ users we can get to the empty state in $M$ steps with non-zero probability and we can go from the empty state to any state sate that has $\bar{M}\in \mathbb{N}$ total users in $\bar{M}$ steps with non-zero probability.

Let $V: \{\mathbb{N}\}^{N_{\epsilon}} \rightarrow \mathbb{R}_{+}$ and define the drift $\Delta V (\zeta)=\mathbb{E}\left[ V(\barbelow{\Phi}^{(d)}_{1})-V(\barbelow{\Phi}^{(d)}_{0}) |\barbelow{\Phi}^{(d)}_{0}=\zeta \right]$, where $\zeta=[k_i]_{i=1}^{N_{\epsilon}} \in \{\mathbb{N}\}^{N_{\epsilon}}$. Then by Foster's Theorem \cite[theorem 5.1.1]{Markov_Pierre13},  if there is a finite set $C$ in the power set of  $\{\mathbb{N}\}^{N_{\epsilon}}$ and $\beta,\alpha>0$ such that:
\begin{align}
\Delta V(\zeta) \leq \beta \mathbbm{1}\{\zeta\in C\} -\alpha \mathbbm{1}\{\zeta\not\in C\}\label{Eq:Suf_drift1st},
\end{align}
then the irreducible and aperiodic  Markov chain  $\barbelow{\Phi}^{(d)}$ is positive recurrent and  ergodic. Let $V(\zeta)=\sum\limits_{i=1}^{N_{\epsilon}} k_i$, so $V$ counts the total number of nodes in $\zeta$. Then,
\begin{align}
\Delta V(\zeta) &=(1)\frac{p}{p+q}+(-1)\frac{q}{p+q}=\frac{p}{p+q}-\frac{q}{p+q}.\label{Eq:Suf_drift2nd} \\
&=\frac{\lambda |\mathcal{D}| -\sum\limits_{i=1}^{N_{\epsilon}}\frac{B\mu}{\ln (2)} \frac{k_i\barbelow{L}^{(\epsilon)}_i}{\sum\limits_{j=1}^{N_{\epsilon}}k_j\bar{L}^{(\epsilon)}_j-\bar{L}^{(\epsilon)}_i+\bS}}{\lambda |\mathcal{D}|+	\sum\limits_{i=1}^{N_{\epsilon}}\frac{B\mu}{\ln (2)} \frac{k_i\barbelow{L}^{(\epsilon)}_i}{\sum\limits_{j=1}^{N_{\epsilon}}k_j\bar{L}^{(\epsilon)}_j-\bar{L}^{(\epsilon)}_i+\bS}}
=
\frac{\frac{\lambda |\mathcal{D}| \ln (2)}{B\mu} -\sum\limits_{i=1}^{N_{\epsilon}} \frac{k_i\barbelow{L}^{(\epsilon)}_i}{\sum\limits_{j=1}^{N_{\epsilon}}k_j\bar{L}^{(\epsilon)}_j-\bar{L}^{(\epsilon)}_i+\bS}}{\frac{\lambda |\mathcal{D}| \ln (2)}{B\mu}+	\sum\limits_{i=1}^{N_{\epsilon}} \frac{k_i\barbelow{L}^{(\epsilon)}_i}{\sum\limits_{j=1}^{N_{\epsilon}}k_j\bar{L}^{(\epsilon)}_j-\bar{L}^{(\epsilon)}_i+\bS}}.
\end{align}

Note that we want to prove that if $\lambda <\frac{B \mu}{\ln(2) |\mathcal{D}|}$, then $\barbelow{\Phi}^{(d)}$ is ergodic. Hence it is enough to show that Foster's theorem is satisfied for $\frac{\lambda |\mathcal{D}| \ln (2)}{B\mu}=1-\delta$ for all $0<\delta<1$.  Let $C=\left\{\zeta: \sum\limits_{i=1}^{N_{\epsilon}}k_i\barbelow{L}^{(\epsilon)}_i \leq M\right\}$, where $M \in \mathbb{R}_{+}$. Then, we have to show that the following is satisfied for some $\alpha>0$ for all $\zeta \notin C$:
\begin{align}
\frac{(1-\delta) - \frac{\sum\limits_{i=1}^{N_{\epsilon}}k_i\barbelow{L}^{(\epsilon)}_i}{\sum\limits_{i=1}^{N_{\epsilon}}k_i\bar{L}^{(\epsilon)}_i+\bS}}{(1-\delta)+	 \frac{\sum\limits_{i=1}^{N_{\epsilon}} k_i\barbelow{L}^{(\epsilon)}_i}{\sum\limits_{i=1}^{N_{\epsilon}}k_i\bar{L}^{(\epsilon)}_i+\bS}} &\leq -\alpha,\\
%\frac{\sum\limits_{i=1}^{N_{\epsilon}}k_i\barbelow{L}^{(\epsilon)}_i}{\sum\limits_{i=1}^{N_{\epsilon}}k_i\bar{L}^{(\epsilon)}_i+\bS}&\geq (1-\delta) \frac{1+\alpha}{1-\alpha},\\
\frac{\sum\limits_{i=1}^{N_{\epsilon}}k_i\barbelow{L}^{(\epsilon)}_i}{\sum\limits_{i=1}^{N_{\epsilon}}k_i\bar{L}^{(\epsilon)}_i+\bS}&\geq (1-\delta) \tilde{\alpha},\label{Eq:Suf_finalForm}
\end{align}
where $\tilde{\alpha}=\frac{1+\alpha}{1-\alpha}$, so that $\tilde{\alpha}$ can be tuned to any value larger than one. Hence, $(1-\delta) \tilde{\alpha}$ can always be tuned to a value strictly less than one. For the LHS, we have the following bounds:
\begin{align}
\frac{\sum\limits_{i=1}^{N_{\epsilon}}k_i\barbelow{L}^{(\epsilon)}_i}{\sum\limits_{i=1}^{N_{\epsilon}}k_i\bar{L}^{(\epsilon)}_i+\bS} \leq \frac{\sum\limits_{i=1}^{N_{\epsilon}}k_i\barbelow{L}^{(\epsilon)}_i}{\sum\limits_{i=1}^{N_{\epsilon}}k_i\bar{L}^{(\epsilon)}_i} \leq 1 \label{Eq:Suf_bound},
\end{align}
where the first inequality follows by neglecting the noise term, and the second follows since $\bar{L}^{(\epsilon)}_i\geq\barbelow{L}^{(\epsilon)}_i, \ \forall i \in \{1, 2, \cdots \}$. Moreover, since we are focusing on the set where $\sum\limits_{i=1}^{N_{\epsilon}}k_i\barbelow{L}^{(\epsilon)}_i > M$, where $M\in \mathbb{R}_{+}$, we can arbitrary approach the second term in \eqref{Eq:Suf_bound} by choosing a larger $M \gg \bS$. Moreover, by reducing $\epsilon$ (increasing $N_{\epsilon}$), we can get arbitrary close to 1 because of \eqref{Eq:Thm1_lim1}. Hence, for very large $M$ and $N_{\epsilon}$, we can write the following:
\begin{align}
\frac{\sum\limits_{i=1}^{N_{\epsilon}}k_i\barbelow{L}^{(\epsilon)}_i}{\sum\limits_{i=1}^{N_{\epsilon}}k_i\bar{L}^{(\epsilon)}_i+\bS}=1-\upsilon(M)-\upsilon(\epsilon),
\end{align}
where $\upsilon(\epsilon)$ and $\upsilon(M)$ can be tuned to any value arbitrary close to zero. Hence, the condition in \eqref{Eq:Suf_finalForm} is satisfied by picking the triple $(M,\epsilon,\tilde{\alpha})$ such that $(1-\delta)\leq\frac{1-\upsilon(M)-\upsilon(\epsilon)}{\tilde{\alpha}}$, which is possible since we can set the triple $(\upsilon(M),\upsilon(\epsilon),\tilde{\alpha})$ to any value that is close to zero.

To complete the proof, we have to show that the drift is bounded by a finite number inside the set $C$. But it is clear from  \eqref{Eq:Suf_drift2nd} that the drift is upper-bounded by $1$. Hence, we can conclude that if $\lambda<\frac{B \mu}{\ln(2) |\mathcal{D}|}$, then $\barbelow{\Phi}^{(d)}$ is ergodic (stable) and the Markov chain admits a unique stationary distribution. 

Note that it is not enough to show that the embedded chain is positive recurrent to deduce that the corresponding CTMC is also positive recurrent with a unique stationary regime \cite[Chapter 7]{Stochastic_Gallager13}. Let $\pi_i$ be the probability that $\barbelow{\Phi}^{(d)}$ is in state $i$ in the stationary regime which we know that is exits and well-defined since $\barbelow{\Phi}^{(d)}$ is irreducible, aperiodic, and positive recurrent. Also, let $v_i$ be the holding-interval parameter which is equal to the sum of all transition rates out of the state $i$ of $\barbelow{\Phi}$, then by Theorem 7.2.6 in \cite{Stochastic_Gallager13}, $\barbelow{\Phi}$ is ergodic and has a unique stationary regime if $\sum_{i} \pi_i/v_i$ is finite. In our case, $v_i$ is lower bounded by the sum of transitions that occur due to an arrival only which is given by \eqref{Eq:arravalRate}, hence, $\sum_{i} \pi_i/v_i\leq\frac{1}{\lambda |\mathcal{D}|} \sum_{i} \pi_i=\frac{1}{\lambda |\mathcal{D}|}$. Hence, $\sum_{i} \pi_i/v_i$ is finite as long as the arrival rate is finite and, by \cite[Theorem 7.2.6]{Stochastic_Gallager13}, we can conclude that $\barbelow{\Phi}$ is ergodic (stable) with a unique stationary regime.

\section{Proof of Theorem 3}\label{App:Thm3}

Similar to the previous proof, we study the embedded chain of $\bar{\Phi}$ denoted by $\bar{\Phi}^{(d)}$. Define the following:
\begin{align}
p_i&=\lambda \epsilon,\\
p&=\sum\limits_{i=1}^{N_{\epsilon}}p_i=\lambda |\mathcal{D}|,\\
q_i&=	\frac{B\mu}{\ln (2)} \frac{k_i\bar{L}^{(\epsilon)}_i}{(k_i-1)\barbelow{L}^{(\epsilon)}_i+\sum\limits_{\substack{j=1\\j\neq i}}^{N_{\epsilon}}k_j\barbelow{L}^{(\epsilon)}_j+\bS},\\
q&=\sum\limits_{i=1}^{N_{\epsilon}}\frac{B\mu}{\ln (2)} \frac{k_i\bar{L}^{(\epsilon)}_i}{(k_i-1)\barbelow{L}^{(\epsilon)}_i+\sum\limits_{\substack{j=1\\j\neq i}}^{N_{\epsilon}}k_j\barbelow{L}^{(\epsilon)}_j+\bS}.
\end{align}

Hence, the transition probabilities for the element $k_i$ are $k_i \rightarrow k_i+1 \ w.p. \ \frac{p_i}{p+q}$ and  $k_i \rightarrow k_i-1 \ w.p. \ \frac{q_i}{p+q}$. It is also clear that $\bar{\Phi}^{(d)}$ is irreducible. To prove that the Markov chain is transient, we cannot use Foster's theorem, but we can use the following theorem \cite[Theorem 2.2.7]{Topics_Fayolle95}: 
\begin{theorem}\label{Thm:transApp}
	For an irreducible Markov chain $\mathcal{L}$ to be transient, it suffices that there exist a positive function $V(\zeta), \zeta \in \{\mathbb{N}\}^{N_{\epsilon}}$, a bounded integer-valued positive function $f(\zeta), \zeta \in \{\mathbb{N}\}^{N_{\epsilon}}$, and numbers $\alpha,M>0$, such that, setting $C=\{\zeta : V(\zeta)\leq M\}$, the following conditions hold:
	\begin{enumerate}
		\item $\sup_{\zeta} f(\zeta)<\infty$
		\item for some $d>0$, the inequality $|V(\zeta_i)-V(\zeta_j)|>d$ implies $p_{ij}=0$.
		\item $\mathbb{E}\left[ V(\mathcal{L}_{f(\mathcal{L}_0)})-V(\mathcal{L}_{0}) |\mathcal{L}_{0}=\zeta)\right]\geq \alpha$ for all $\zeta \notin C$.
	\end{enumerate}
\end{theorem}

To use this theorem,  let $V(\zeta)=\sum\limits_{i=1}^{N_{\epsilon}} k_i$, $f(\cdot)=1$, and $d=1$. Then it is clear that the first two conditions are satisfied, since the chain can only jump to states that has one more or one less node in it. For the third condition, we can write the following with a bit of algebra: 

\begin{align}
\sum\limits_{i=1}^{N_{\epsilon}}	 \frac{k_i\bar{{L}}^{(\epsilon)}_i}{(k_i-1)\barbelow{L}^{(\epsilon)}_i+\sum\limits_{\substack{j=1\\j\neq i}}^{N_{\epsilon}}k_j\barbelow{L}^{(\epsilon)}_j+\bS}&\leq \frac{\lambda |\mathcal{D}| \ln (2)}{B\mu} \tilde{\alpha},\label{Eq:Nes_finalForm}
\end{align}
where $\tilde{\alpha}=\frac{1- \alpha}{1+\alpha}$, so $\tilde{\alpha}$ can be tuned to any positive value less than one. Since we want to show that the Markov chain is transient for all $\frac{\lambda |\mathcal{D}| \ln (2)}{B\mu}>1$, it is enough to show it for $\frac{\lambda |\mathcal{D}| \ln (2)}{B\mu}=1+\delta$, where $\delta$ is a strictly positive number. Hence, the RHS $(1+\delta) \tilde{\alpha}$ can always be tuned to a value strictly larger than one by choosing the appropriate $\tilde{\alpha}$ . For the LHS, one can find the following:
\begin{align}
\sum\limits_{i=1}^{N_{\epsilon}}	 \frac{k_i\bar{{L}}^{(\epsilon)}_i}{(k_i-1)\barbelow{L}^{(\epsilon)}_i+\sum\limits_{\substack{j=1\\j\neq i}}^{N_{\epsilon}}k_j\barbelow{L}^{(\epsilon)}_j+\bS}&\leq\sum\limits_{i=1}^{N_{\epsilon}}	 \frac{k_i\bar{{L}}^{(\epsilon)}_i}{(k_i-1)\barbelow{L}^{(\epsilon)}_i+\sum\limits_{\substack{j=1\\j\neq i}}^{N_{\epsilon}}k_j\barbelow{L}^{(\epsilon)}_j},\\
&\leq\sum\limits_{i=1}^{N_{\epsilon}}	 \frac{k_i\bar{{L}}^{(\epsilon)}_i}{\sum\limits_{j=1}^{N_{\epsilon}}(k_j-1)\barbelow{L}^{(\epsilon)}_j\mathbbm{1}\{k_j>0\}}, \\
&= \frac{\sum\limits_{i=1}^{N_{\epsilon}}	k_i\bar{{L}}^{(\epsilon)}_i}{\sum\limits_{i=1}^{N_{\epsilon}}k_i\barbelow{{L}}^{(\epsilon)}_i-	 \sum\limits_{i=1}^{N_{\epsilon}}\barbelow{L}^{(\epsilon)}_i\mathbbm{1}\{k_i>0\}} \label{Eq:Nes_q12}.
\end{align}

Note that $\sum\limits_{i=1}^{N_{\epsilon}}\barbelow{L}^{(\epsilon)}_i\mathbbm{1}\{k_i>0\} \leq L_{\rm max} N_{\epsilon}$ also $\sum\limits_{i=1}^{N_{\epsilon}}k_i\barbelow{{L}}^{(\epsilon)}_i\geq L_{\rm min} M$ since we are focusing on states outside the set $C$. Hence, for a fixed $N_{\epsilon}$, we can choose $M \gg \frac{L_{\rm max}N_{\epsilon}}{L{\rm min}}$ such that the term in \eqref{Eq:Nes_q12} can be written as:
\begin{align}
\frac{\sum\limits_{i=1}^{N_{\epsilon}}	k_i\bar{{L}}^{(\epsilon)}_i}{\sum\limits_{i=1}^{N_{\epsilon}}k_i\barbelow{{L}}^{(\epsilon)}_i} \label{Eq:Nes_q13} +\upsilon_1(N_{\epsilon},M),
\end{align}   
where $\upsilon_1(N_{\epsilon},M) \ll 1$. Moreover, let $\beta=\max_{j}\frac{k_j \bar{L}^{(\epsilon)}_j}{k_j \barbelow{L}^{(\epsilon)}_j}=\max_{j}\frac{\bar{L}^{(\epsilon)}_j}{\barbelow{L}^{(\epsilon)}_j}$, then $\frac{k_i \bar{L}^{(\epsilon)}_i}{k_i \barbelow{L}^{(\epsilon)}_i} \leq \beta$ for all $i \in \{1, 2, \cdots, N_{\epsilon}\}$, which leads to
\begin{align}
1\leq \frac{\sum\limits_{i=1}^{N_{\epsilon}}	k_i\bar{{L}}^{(\epsilon)}_i}{\sum\limits_{i=1}^{N_{\epsilon}}k_i\barbelow{{L}}^{(\epsilon)}_i} \leq \max_{j}\frac{\bar{L}^{(\epsilon)}_j}{\barbelow{L}^{(\epsilon)}_j}.
\end{align}

Hence, by increasing $N_{\epsilon}$, this term can get arbitrary close to $1$ due to \eqref{Eq:Thm1_lim1}. Hence, for large $N_{\epsilon}$ and larger $M \gg \frac{L_{\rm max}N_{\epsilon}}{L{\rm min}}$, we can write the LHS in \eqref{Eq:Nes_finalForm} as $1+\upsilon_2(N_{\epsilon},M)+\upsilon_2(N_{\epsilon})$, where $\upsilon_2(N_{\epsilon})$ can be set to any positive value very close to zero. Overall, we can rewrite the condition in  \eqref{Eq:Nes_finalForm} for large $N_{\epsilon}$ and larger $M \gg \frac{L_{\rm max}N_{\epsilon}}{L{\rm min}}$ as
\begin{align}
1+\upsilon_2(N_{\epsilon},M)+\upsilon_2(N_{\epsilon}) \leq (1+\delta)\tilde{\alpha}.
\end{align}

Hence we can always set $\tilde{\alpha}$ to a small value such that $(1+\delta)\tilde{\alpha}$ is a fixed value $\tilde{\alpha}_{2}$ strictly larger than one and then set $N_{\epsilon}$ and $M$ to large values such that the LHS is less than $\tilde{\alpha}_{2}$. Hence, for all $\frac{\lambda |\mathcal{D}| \ln (2)}{B\mu}>1$, the conditions in Theorem \ref{Thm:transApp} are satisfied and $\bar{\Phi}^{(d)}$ is transient. Finally, since the embedded chain of $\bar{\Phi}$ is transient, we can conclude that $\bar{\Phi}$ is also transient which completes the proof.

\section{Mean-field Model}\label{App:MeanF}
In this section, our aim is to describe a more general model that captures our network model as a special case, but asymptotically behaves exactly as the mean-field model we discussed, i.e., the first order approximation. Hence, it bridges the gap between the true network performance and the first order approximation. We focus on the discrete model for simplicity. Let $N_f$ be the number of frequency bands utilized by the network. At each time step, users arrive to the network at a rate $N_f \lambda$ and each user operates at a frequency band chosen uniformly at random. Moreover, at the beginning of each time step, current nodes in the network randomly and uniformly shuffle between the frequency bands. Clearly, this setup simplifies to the original model we discussed in this paper when $N_f$ is set to one. Let $F_{i,t}$ be a random variable that represents the frequency band chosen by the $i^{\rm th}$ node at time step $t$. Then the data rate this node gets at time $t$ is
\begin{align}
 \frac{B}{\ln(2)}  \frac{L(x_i)^{1-l}}{\sum\limits_{j\in \Phi_t\setminus\{i\}}L(x_j)^{1-l} \mathbbm{1} \{ f_i=f_j\}+\bS} , \ \ \ \ \ \forall i  \in \Phi_t,\label{Eq:RateConsPriMean}
\end{align}
where $\Phi_t=\{1,2, \cdots \}$ is the a numbering for active nodes in the network at time $t$, $x_i$ is the location of the $i^{\rm th}$ node and $f_i$ is the frequency band chosen by this node.

\begin{claim}
As $N_f \rightarrow \infty$, the system described above behaves exactly as the mean-field approximation in the steady state.
\end{claim}

The intuition behind this claim is as follows. As $N_f$ gets larger, the correlation between the active nodes operating at the same frequency band dissolves, since the interfering nodes could have been operating on any of the $N_f$ frequency bands in the previous time steps, and there are infinitely many of them. Hence, the presence of a node at location $x$ does not tell us anything about the location of the interfering nodes, since they were operating at a different frequency band at the previous time step with a very high probability. Hence, the correlation, which caused the gap between the true system performance $N_f=1$ and the first order approximation $N_f\rightarrow \infty$ disappears due to the shuffling process that occurs at the beginning of each time step. Proving the previous claim is of great interest as a theoretical problem, however, we postpone it to future work, since it does not add much value to the current work, especially given that it requires a substantial amount of work.

\section{Proof of Theorem 4}\label{App:FOA}
By assuming that $\Phi$ is a PPP with intensity function $\gamma_p(\cdot)$, \eqref{Eq:RateConsPri} can be evaluated as follows:
\begin{align}\label{Eq:SC_2_1}
\rho&=\gamma_p(x) \frac{B}{\ln(2)} \mathbb{E}\left[ \frac{L(x)^{1-l}}{\sum\limits_{y\in \Phi \setminus\{x\}}L(y)^{1-l}+\bS}\right], \ \ \ \ \ \forall x  \in \mathcal{D},\\
&= \frac{B}{\ln(2)} \gamma_p(x) L(x)^{1-l} \int\limits_{0}^{\infty} e^{-t \bS} \exp \left(- \int\limits_{\mathcal{D}} \left( 1- e^{-t L(y)^{1-l}}\right) \gamma_p(y) \mathrm{d}y \right) \mathrm{d}t, \label{Eq:SC_3}
\end{align}
where \eqref{Eq:SC_2_1} follows due to the independence property of the PPP and \eqref{Eq:SC_3} follows because, for a positive random variable $Z$ with a probability density function (pdf) $f_{Z}(\cdot)$, $\mathbb{E}[Z^{-1}]=\int\limits_{0}^{\infty} \mathcal{L}_{Z}(t)\mathrm{d}t$, where $\mathcal{L}_{Z}(\cdot)$  is the Laplace transform of $f_{Z}(\cdot)$. Then by using the probability generating functional of a PPP \cite{Stochastic_Baccelli10} with intensity $\gamma_p(\cdot)$ we get \eqref{Eq:SC_3}. Define
\begin{align}
G(\gamma_p)= \int\limits_{0}^{\infty} e^{-t \bS} \exp \left(- \int\limits_{\mathcal{D}} \left( 1- e^{-t L(y)^{1-l}}\right) \gamma_p(y) \mathrm{d}y \right) \mathrm{d}t, \label{Eq:SC_4}
\end{align}
which is a function of the intensity function $\gamma_p(\cdot)$ only and independent of $x$. Then
\begin{align}
\frac{\rho \ln(2)}{B} &= \gamma_p(x)  L(x)^{1-l} G(\gamma_p), \ \ \ \ \ \forall x  \in \mathcal{D},\\
\gamma_p(x)&=\frac{\rho \ln(2)}{B  L(x)^{1-l} G(\gamma_p)}, \ \ \ \ \ \forall x  \in \mathcal{D}. \label{Eq:SC_5}
\end{align}

Note that $\frac{\rho \ln(2)}{B G(\gamma_p)}$ is independent of $x$. Hence, $\gamma_p(x)$ is inversely proportional to the path loss function after inversion $L(x)^{1-l}$. By substituting \eqref{Eq:SC_5} in \eqref{Eq:SC_4}, we get
\begin{align}
G(\gamma_p)= \int\limits_{0}^{\infty} e^{-t \bS} \exp \left(-\frac{\rho \ln(2)}{B G(\gamma_p)} \int\limits_{\mathcal{D}} \left( 1- e^{-t L(y)^{1-l}}\right) L(y)^{l-1} \mathrm{d}y \right) \mathrm{d}t, \label{Eq:SC_6}
\end{align}
which is a fixed point equation in $G(\gamma_p)$ since $\rho$ is known. The form given in the theorem is found by a simple change of variables.

\section{Proof of Corollary \ref{Cor:SolutionsFOA}}\label{App:proofSolutionsFOA}

   Let $\bar{N}=\gamma |\mathcal{D}|$ which is the average number of users in the cell under the first order approximation and define $C:=\frac{\rho \ln(2) |\mathcal{D}|}{B}$ and $f(\bar{N}):=\bar{N} e^{-\bar{N} } \int\limits_{0}^{\infty}  \exp \left( -t \bS +\bar{N}  e^{-t }  \right) \mathrm{d}t$. Then the fixed point equation can be written as 
   \begin{align}
       C&=f(\bar{N})\\
       &=\bar{N} \int\limits_{0}^{\infty}  \exp \left( -t \bS -\bar{N}(1-  e^{-t } ) \right) \mathrm{d}t\label{Eq:FOA_F1}\\
       &=\bar{N}^{1-\bS} e^{-\bar{N} } \int\limits_{0}^{\bar{N}}  e^{y} y^{\bS-1}\mathrm{d}y \label{Eq:FOA_F2}\\
	&=\frac{\bar{N} e^{-\bar{N}}}{\bS} {}_{1}F_1(\bS,\bS+1,\bar{N}),\label{Eq:FOA_F3}
   \end{align}
   	where \begin{align}\label{Eq:HyperGeoInt}
   	{}_{1}F_1(a,b,z)&=\frac{\Gamma(b)}{\Gamma(a)\Gamma(b-a)}\int\limits_{0}^{1} e^{zt}t^{a-1} (1-t)^{b-a-1}\mathrm{d}t, \ a,b,z>0\\
   	&=\sum\limits_{j=0}^{\infty} \frac{(a)_j}{(b)_j} \frac{z^{j}}{j!}, b>0,
   	\label{Eq:HacyperGeoSer}
   	\end{align}
   	is the confluent Hypergeometric function\footnote{Note that the integral representation is only valid when $a\geq0$, while the series representation is for a general $a$. Also, note that in some texts, a different definition is used which does not have the term $\Gamma(b)$, as in \cite[Eq. 13.2.3]{NIST_Olver10}, while the one used in this work is \cite[Eq. 13.2.2]{NIST_Olver10}.} of the first kind \cite[Chapter 13]{NIST_Olver10} and $\Gamma(\cdot)$ is the complete gamma function \cite[Eq. 5.2.1]{NIST_Olver10}. Note that \eqref{Eq:FOA_F2} is found by the substitution $y=\bar{N}e^{-t}$ and \eqref{Eq:FOA_F3} by substituting $z=\frac{y}{\bar{N}}$ in~\eqref{Eq:FOA_F2}. 
   	
		   For the first point in the corollary, it is clear from \eqref{Eq:FOA_F1} that $f(0)=0$. Moreover, based on \cite[Eq. 13.2.23]{NIST_Olver10}, $\frac{\bar{N}e^{-\bar{N}}}{\bS}{}_{1}F_1(\bS,\bS+1,\bar{N})\rightarrow 1$ as $\bar{N} \rightarrow \infty$, hence, $\lim\limits_{\bar{N}\rightarrow \infty}f(\bar{N})=1$.

		    For the case of $\bS \geq 1$, we will show that $f(\bar{N})$ is strictly increasing in $\bar{N}$, and given that $\lim\limits_{\bar{N}\rightarrow \infty}f(\bar{N})=1$, we prove the desired statement. To this end, we prove that the first derivative is strictly positive for any finite $\bar{N}$. The following lemma will be used in the proof.
		    \begin{lemma}
		    The following are properties of the confluent Hypergeometric function assuming $0<a,b<\infty$ and $0\leq z<\infty$.
            \begin{align}
                &{}_1 F_{1}(a,a+1,z)\geq1,
                \label{eq:Hyp_1}\\
                &\frac{d\ {}_1 F_{1}(a,b,z)}{dz}=\frac{a}{b} {}_1 F_{1}(a+1,b+1,z),\label{eq:Hyp_2}\\
                &{}_1 F_{1}(a+1,a+2,z)=\frac{(a+1)(z-a){}_1 F_{1}(a,a+1,z)}{az} +\frac{(a+1){}_1 F_{1}(a-1,a,z)}{z} ,  \label{eq:Hyp_3}\\
               &{}_1 F_{1}(a,a+1,z) \geq  \frac{a}{z} \left(e^{z}-1\right), \  \forall \ a\leq1.\label{eq:Hyp_5}
            \end{align}
            \begin{proof}
            Equation \eqref{eq:Hyp_1} follows  from the definition in \eqref{Eq:HyperGeoInt} and  noting that $e^{t \bar{N}}\geq 1$ for all $t\in[0,1]$ and for all $\bar{N} \geq0$. \eqref{eq:Hyp_2} is taken from \cite[Eq. 13.3.15	]{NIST_Olver10}. \eqref{eq:Hyp_3} is taken from \cite[Eq. 16.1.9c]{Handbook_Cuyt09}. Finally, \eqref{eq:Hyp_5} follows from the definition in \eqref{Eq:HyperGeoInt} and  noting that $t^{a-1}\geq 1$ for $t\in[0,1]$ for all $a\leq1$.
            \end{proof}
		    \end{lemma}
		    
		    Using the results from the previous lemma, the first derivative can be written as,
		    \begin{align}
		   \frac{df(\bar{N})}{d\bar{N}}&=
		  \frac{e^{-\bar{N}}}{\bS} \left( \frac{\bS\bar{N}}{\bS+1} {}_1 F_{1}(\bS+1,\bS+2,\bar{N})+(1-\bar{N}) {}_1 F_{1}(\bS,\bS+1,\bar{N})\right)\label{Eq:FOA_Der}\\
 &=\frac{e^{-\bar{N}}}{\bS}\left((1-\bS){}_1 F_1(\bS,\bS+1,\bar{N})+\bS {}_1 F_1(\bS-1,\bS,\bar{N}) \right)\label{Eq:FOA_Der2}\\
 &=\frac{e^{-\bar{N}}}{\bS}\left((1-\bS) \sum\limits_{j=0}^{\infty} \frac{\bS}{\bS+j} \frac{\bar{N}^{j}}{j!}+\bS \sum\limits_{j=0}^{\infty} \frac{(\bS-1)}{\bS+j-1} \frac{\bar{N}^{j}}{j!} \right)\label{Eq:FOA_Der3}\\
      &=(\bS-1)e^{-\bar{N}}\left( \sum\limits_{j=0}^{\infty} \frac{1}{(\bS+j-1)(\bS+j)} \frac{\bar{N}^{j}}{j!}\right)\label{Eq:FOA_Der4},
		    \end{align}
		    where \eqref{Eq:FOA_Der} follows from \eqref{eq:Hyp_2}, \eqref{Eq:FOA_Der2} follows from \eqref{eq:Hyp_3}, \eqref{Eq:FOA_Der2} follows from expressing  ${}_1 F_1(\cdot,\cdot,\cdot)$ in its series representation, and \eqref{Eq:FOA_Der4} follows from simple mathematical manipulations. Note that for the case of $\bS>1$, $(\bS-1)$ is positive and all the terms of the series are strictly  positive as well. Hence, adding $\lim\limits_{\bar{N} \rightarrow \infty}f(\bar{N})=1$ to the picture, we know the fixed point equation has a unique solution if and only if $C\in[0,1]$ and no solution if $C>1$.
		   
		    \begin{lemma}
		    For all $\bS<1$, we have the following.
		    \begin{enumerate}[{2}.1:]
		        \item $f( \bar{N})\geq(1-e^{- \bar{N}})$ and $f( \bar{N})\geq\frac{ \bar{N} e^{-\bar{N}}}{\bS}$ for all $ \bar{N}\geq0$.
		        \item $f(\bar{N})$ is strictly increasing for $0\leq\bar{N}\leq1$.
		        \item $f(\bar{N})\leq \frac{2-b}{1-b} b^{\bS-1}$ for any $b\in(0,1)$ and for all $\bar{N}\geq0$
		        \item $f(\bar{N})$ is strictly decreasing for $\frac{\bS+1}{\bS-1}\leq\bar{N}<\infty$.
		        \item $f(\bar{N})$ has a single maximum that happens when $1<\bar{N}<\frac{\bS+1}{\bS-1}$.
		    \end{enumerate}
		    \begin{proof}
		    The first lower bound follows directly from \eqref{eq:Hyp_5} and the second follows from \eqref{eq:Hyp_1}. $2.2$ is clear from the first derivative expression in \eqref{Eq:FOA_Der}, since all terms are positive if $\bar{N}\leq1$. To prove $2.3$, let $b=(0,1)$, then
\begin{align}
    f(\bar{N})= \bar{N} \int\limits_{0}^{1} e^{-\bar{N}(1-t)} t^{\bS-1} \mathrm{d}t= \bar{N} \int\limits_{0}^{b} e^{-\bar{N}(1-t)} t^{\bS-1} \mathrm{d}t +\bar{N} \int\limits_{b}^{1} e^{-\bar{N}(1-t)} t^{\bS-1} \mathrm{d}t.
\end{align}
But,
\begin{align}
  \bar{N} \int\limits_{b}^{1} e^{-\bar{N}(1-t)} t^{\bS-1} \mathrm{d}t &\leq b^{\bS-1}  (1-e^{(b-1)\bar{N}})\leq b^{\bS-1}, \label{eq:F_Bound_1}\\
  \bar{N} \int\limits_{0}^{b} e^{-\bar{N}(1-t)} t^{\bS-1} \mathrm{d}t &\leq\bar{N} e^{-\bar{N}(1-b)} \frac{b^{\bS}}{\bS} \leq \frac{b^{\bS}e^{-1}}{(1-b)\bS} \leq\frac{b^{\bS}}{(1-b)\bS} ,\label{eq:F_Bound_2}
\end{align}
where the first inequality in \eqref{eq:F_Bound_1} follows by replacing $t^{\bS-1}$ by $b^{\bS-1}$ and evaluating the integral, and the second by noting that $(1-e^{(b-1)\bar{N}})\leq 1$. 
The first inequality in \eqref{eq:F_Bound_2} follows by substituting $t$ in $e^{-\bar{N}(1-t)}$ by $b$ and evaluating the integral, the second follows from the fact that $\bar{N} e^{-\bar{N}(1-b)}$ is log-concave w.r.t. $\bar{N}$ and its maximum is $\frac{e^{-1}}{1-b}$, and the third by replacing $e^{-1}$ by 1. Hence, $f(\bar{N})\leq b^{\bS-1}\left( \frac{b}{(1-b)\bS} +1\right)$ and since it holds for any $b \in (0,1)$, we choose $b=0.5$ in the original statement in the corollary.

For the fourth point, we need to rewrite the first derivative in \eqref{Eq:FOA_Der4} as
	\begin{align}
	    (1-\bS)e^{-\bar{N}}\left( \frac{1}{(1-\bS)\bS}+\frac{-\bar{N}}{(\bS+1)\bS}  +  \sum\limits_{j=2}^{\infty} \frac{-1}{(\bS+j-1)(\bS+j)} \frac{\bar{N}^{j}}{j!}\right).
	\end{align}	    
	
	Note that $(1-\bS)>0$ and $\frac{1}{(1-\bS)\bS}$ is the only positive term in the summation. Hence, it is sufficient to pick a $\bar{N}$ such that  $ \frac{1}{(1-\bS)\bS}+\frac{-\bar{N}}{(\bS+1)\bS}$ is negative to prove that the first derivative is negative, which is satisfied if $\bar{N}\geq \frac{1+\bS}{1-\bS}$. Moreover, $\frac{1}{(1-\bS)\bS}$ is independent of $\bar{N}$ and $\frac{\bar{N}}{(\bS+1)\bS}  +  \sum\limits_{s=2}^{\infty} \frac{1}{(\bS+s-1)(\bS+s)} \frac{\bar{N}^{s}}{s!}$ is strictly increasing in $\bar{N}$, which means that the first derivative could switch its sign only once, and hence $f(\bar{N})$ has a single maximum that occurs between $1<\bar{N}<\frac{1+\bS}{1-\bS}$, since it is increasing before $\bar{N}=1$ and decreasing after $\bar{N}=\frac{1+\bS}{1-\bS}$.
   \end{proof}
		    \end{lemma}	
		    
		    Using the results from the previous lemma, it is straightforward to deduce the statements made in the corollary.

		    \section{Proof of Theorem \ref{Th:SOA}}\label{App:SOA}
    As in Section \ref{Sec:StabCond}, we divide the region $\mathcal{D}$ into $N_{\epsilon}$ disjoint connected sets $A^{(\epsilon)}_{j}, j\in \{ 1, 2, \cdots, N_{\epsilon}\}$ with equal areas $ \epsilon=\frac{|\mathcal{D}|}{N_{\epsilon}}$. Furthermore, the center of the region $A^{(\epsilon)}_{j}$ is denoted by $x_j$. Let $N_{\epsilon}$ be large enough for the probability that more that  one user exist in any region in the stationary case to be negligible. Also, let us focus on a tiny period of time given by $\epsilon_t$, such that the probability of the arrival of more than one user within $\epsilon_t$ is negligible which is possible due to the Poisson arrivals.

%Let's discretize the region  into $n^2$ small layers, each with area $ \pi (\frac{a}{2 n})^2$ and inner radius $r_m, m \in\{1,..,n^2\}$ as shown in Fig. \ref{fig:CirRegion}. Hence, $r_m=\frac{a}{2n}\sqrt{m} $.  Furthermore, let $n$ be large enough such that at maximum one user can exist in any layer in the stationary case. Such an event can be guaranteed with probability one in the stationary case, otherwise we will have an infinite accumulation of users. Also, let us focus on a tiny period of time given by $\epsilon_t$, such that at maximum one user can arrive within this time slot. This event can also be guaranteed with probability one due to the Poisson arrivals.

Let $N_{x_m}$ be the number of users in the region $A^{(\epsilon)}_m$ which is centered at $x_m$. Then, by conditioning on the positions of the other users in the cell, the number of users in a given region can be represented by a two-state Markov chain: ($N_{x_m}=1$) and ($N_{x_m}=0$). The transition rate from ($N_{x_m}=0$) to ($N_{x_m}=1$) is the same as the arrival rate to the region, which is $ \rho\epsilon \epsilon_t$ since the arrival process is Poisson. On the other hand, the transition rate from ($N_{r_m}=1$) to ($N_{r_m}=0$) is the rate a user within the region $A^{(\epsilon)}_m$ gets conditioning on the location of the other users. It is given by $\frac{B}{\ln(2)} \frac{L(x_m)^{1-l}}{I_{x_m}+\bS}\epsilon_t$, where $I_{r_m}$ is the interference from the other nodes in the cell:  $I_{x_m}= \sum_{k=1,k \neq m}^{n^2} L(x_k) \zeta_k$, where $\zeta_k=1$ if a user is located within the region $A^{(\epsilon)}_k$ and zero otherwise. Hence, by the balance equation, we get
\begin{align} \label{Eq:SOH_1}
\rho\epsilon \epsilon_t \mathbb{P}\left( N_{x_m}=0\right)=\mathbb{P}\left( N_{x_m}=1\right)\frac{B}{\ln(2)} \frac{L(x_m)^{1-l}}{I_{x_m}+\bS}\epsilon_t,
\end{align} 
where the equality holds in the limit of $\epsilon,\epsilon_t\rightarrow0$, i.e., there are extra terms that vanish the limit. Since $\gamma^{(1)} (\cdot)$ is the first moment measure of $\Phi$, $\mathbb{P}\left( N_{x_m}=1\right) \approx\epsilon \gamma^{(1)} (x_{m})$ and \eqref{Eq:SOH_1} simplifies to
\begin{align} \label{Eq:SOH_2}
\rho \left(1-\epsilon \gamma^{(1)} (x_{m}) \right)&=\gamma^{(1)} (x_m)\frac{B}{\ln(2)} \frac{L(x_m)^{1-l}}{I_{x_m}+\bS}.
\end{align} 

 Note that the interference is conditional on the random variables  $\zeta_k, k\in{\{1, \cdots,N_\epsilon\}} \backslash m$. Ideally, we want to average both sides in \eqref{Eq:SOH_2} over the random variables $\zeta_k$. However, such an averaging step results in the same rate-conservation equation as in \eqref{Eq:RateConsPri}. Hence, we start by finding the mean of the interference term alone.
\begin{align}\label{Eq:SOH_4}
\mathbb{E} \left[I_{x_{m}}\right]=\mathbb{E} \left[ \sum_{k=1,k \neq m}^{N_\epsilon} L(x_k)^{1-l} \zeta_k \Big|\zeta_m=1 \right]=\sum_{k=1,k \neq m}^{N_\epsilon} L(x_k)^{1-l}  \mathbb{E} \left[ \zeta_k|\zeta_m=1 \right].
\end{align}

%Hence, we assume that the interference seen by the BS is the mean. This can be justified by the fluid limit since we are interested in the high arrival rate case. Hence, we substitute $I_{r_{m}}$ by $\mathbb{E} \left[I_{r_{m}}\right]$, where the expectation  is w.r.t $\zeta_k,\  \forall k \in \{1,..,n^2 \} \backslash m$ given that $\zeta_m=1$.  Hence,

Also,  $\mathbb{E} \left[ \zeta_k|\zeta_m=1 \right]=\mathbb{P} \left\{  \zeta_k=1|\zeta_m=1\right\}=\frac{\mathbb{P} \left\{  \zeta_k=1,\zeta_m=1\right\}}{\mathbb{P} \left\{  \zeta_m=1\right\}}$. Let $\gamma^{(2)} (x_{k},x_{v})$ denote the second moment measure of $\Phi$ as we defined in \eqref{Eq:SecOrdMea}. Hence, $\mathbb{P} \left\{  \zeta_k=1,\zeta_m=1\right\} \approx \gamma^{(2)}(x_k,x_m) \epsilon^2$ and we have already mentioned that $\mathbb{P} \left\{  \zeta_m=1\right\}\approx\epsilon \gamma^{(1)}(x_m)$. Based on these approximations, \eqref{Eq:SOH_4} simplifies to 
\begin{align}\label{Eq:SOH_5}
\mathbb{E} \left[I_{x_{m}}\right]&=\sum_{k=1,k \neq m}^{N_\epsilon} L(x_k)^{1-l}  \mathbb{E} \left[ \zeta_k|\zeta_m=1 \right]=\sum_{k=1,k \neq m}^{N_\epsilon} L(x_k)^{1-l}\frac{\gamma^{(2)} (x_k,x_m)}{\gamma^{(1)}(x_m)} \epsilon,\\
&=-\frac{\gamma^{(2)}(x_m,x_m)}{\gamma^{(1)}(x_m)}L(x_m)^{1-l}\epsilon +\sum_{k=1}^{N_\epsilon} L(x_k)^{1-l}\frac{\gamma^{(2)} (x_k,x_m)}{\gamma^{(1)}(x_m)} \epsilon,\\
&=-\frac{\gamma^{(2)}(x_m,x_m)}{\gamma^{(1)}(x_m)}L(x_m)^{1-l}\frac{|\mathcal{D}|}{N_\epsilon} +\sum_{k=1}^{N_\epsilon} L(x_k)^{1-l}\frac{\gamma^{(2)} (x_k,x_m)}{\gamma^{(1)}(x_m)} \frac{|\mathcal{D}|}{N_\epsilon},\label{Eq:SOH_5_1}\\
\lim_{N_{\epsilon} \rightarrow \infty}\mathbb{E} \left[I_{x_{m}}\right]&= \int\limits_{\mathcal{D}}\frac{\gamma^{(2)} (x,x_m)}{\gamma^{(1)}(x_m)} L(x)^{1-l} \mathrm{d}x, \label{Eq:SOH_6}
\end{align}
where \eqref{Eq:SOH_5_1} follows by substituting the value of $\epsilon$ and \eqref{Eq:SOH_6} follows by taking the limit when $N_{\epsilon}\rightarrow \infty$.

%Hence, \eqref{Eq:SOH_3}  reduces to the following, after substituting $I_{r_m}$ by $\mathbb{E} \left[I_{r_{m}}\right]$ and taking the limit when $n\rightarrow \infty$.
%\begin{align}
%\rho &= \gamma^{(1)} (r)\frac{B}{\ln(2)} \frac{L(r)}{2 \pi \int\limits_{0}^{\frac{a}{2}}uL(u)\frac{\gamma^{(2)} (r,u)}{\gamma^{(1)}(r)}  du+\bS}, \ \ \ \forall r\in \left[0,\frac{a}{2}\right].\label{Eq:SOH_7}
%\end{align}

Hence, to find the distribution of the interference term, we must at least know the second order moment measure. Let us now take the two regions centered at $x_m$ and $x_k$, $m \neq k$. Similar to the previous approach, the joint number of users in the regions with centers $x_m$ and $x_k$ conditional on the location of the other users in the cell can be modeled as a four-state Markov chain, with the states $(0,0),(0,1),(1,0),(1,1)$, where $(a,b)$ is the state $N_{x_m}=a$ and $N_{x_k}=b$. The transition rate from $(0,1)$ to $(1,1)$ is $\rho \epsilon \epsilon_t$ which is also the transition rate from $(1,0)$ to $(1,1)$. The transition rate from $(1,1)$ to $(0,1)$ is the rate a user at $x_m $ gets, which is $\frac{B}{\ln(2)} \frac{L(x_m)^{1-l}}{I_{x_m,x_k}+L(x_k)^{1-l}+\bS}\epsilon_t$, where $I_{x_m,x_k}$ is the interference from all the users in the cell except the user at  $x_m$ and the user at $x_k$:  $I_{x_m,x_k}= \sum_{u=1,u \not\in \{m,k\}}^{N_{\epsilon}} L(x_u)^{1-l} \zeta_u$, where $\zeta_u=1$ if a user exits in the region with center $x_u$. Similarly, the transition rate from $(1,1)$ to $(1,0)$ is $\frac{B}{\ln(2)} \frac{L(x_k)^{1-l}}{I_{x_m,x_k}+L(x_m)^{1-l}+\bS}\epsilon_t$. Note that we cannot jump from the state $(1,1)$ to $(0,0)$ directly, since it requires two events and this happens with a probability that approaches zero. Similarly, we cannot jump directly from $(0,0)$ to $(1,1)$. Hence, using the balance equation, we get the following equation.
% \begin{align} 
% \rho& \epsilon\epsilon_t \mathbb{P}\left( N_{x_m}=1,N_{x_k}=0\right)+\epsilon \epsilon_t \mathbb{P}\left( N_{x_m}=0,N_{x_k}=1\right)=\notag\\
% & \mathbb{P}\left( N_{x_m}=1,N_{x_k}=1\right)\frac{B}{\ln(2)} \left(  \frac{L(x_m)^{1-l}}{I_{x_m,x_k}+L(x_k)^{1-l}+\bS}+ \frac{L(x_k)^{1-l}}{I_{x_m,x_k}+L(x_m)^{1-l}+\bS}\right)\epsilon_t, 
% \end{align} 
\begin{align} \label{Eq:SOH_8}
\rho& \epsilon  \left(\mathbb{P}\left( N_{x_m}=1,N_{x_k}=0\right)+\mathbb{P}\left( N_{x_m}=0,N_{x_k}=1\right) \right)=\notag\\
& \mathbb{P}\left( N_{x_m}=1,N_{x_k}=1\right)\frac{B}{\ln(2)} \left(  \frac{L(x_m)^{1-l}}{I_{x_m,x_k}+L(x_k)^{1-l}+\bS}+ \frac{L(x_k)^{1-l}}{I_{x_m,x_k}+L(x_m)^{1-l}+\bS}\right).
\end{align} 

But $\mathbb{P}\left( N_{x_m}=1,N_{x_k}=1\right) \approx \gamma^{(2)}(x_m,x_k)\epsilon^2$. Hence, the mean of the interference term is
\begin{align}
 \mathbb{E}\left[I_{x_k,x_m}\right]&=  \mathbb{E}\left[\sum_{u=1,u \not\in \{m,k\}}^{N_\epsilon} L(x_u)^{1-l} \zeta_u\Bigg|\zeta_k=1 \&   \zeta_m=1\right]
  \end{align}
  and
 \begin{align}
 \lim_{N_\epsilon \rightarrow \infty} \mathbb{E}\left[I_{x_k,x_m}\right]&= \int\limits_{\mathcal{D}}\frac{\gamma^{(3)}(x_m,x_k,x)}{\gamma^{(2)}(x_m,x_k)} L(x)^{1-l}\mathrm{d}x \label{Eq:SOH_8_1},
\end{align}
where $\gamma^{(3)}(\cdot,\cdot,\cdot)$ is the third moment measure \cite{Stochastic_Baccelli10}. Hence, to evaluate the distribution of $I_{x_k,x_m}$, we need to know at least $\gamma^{(3)}(\cdot,\cdot,\cdot)$. This chain of dependence keeps going on and on as observed in \cite{Can_Baccelli13} in a different context. The reason is that to fully capture the correlations between the users locations, we should account for all different combinations of the existence of users in different regions. Hence, at this point, we have to use some approximations to evaluate the steady-state regime.
\begin{assumption}\label{As:Probs}
Since the joint probabilities $\mathbb{P}\left( N_{x_m}=1,N_{x_k}=0\right)$ and $\mathbb{P}\left( N_{x_m}=0,N_{x_k}=1\right)$ cannot be directly expressed in terms of the moment measures, we factorize them as
\begin{equation}
    \mathbb{P}\left( N_{x_m}=1,N_{x_k}=0\right)=\mathbb{P}\left( N_{x_m}=0,N_{x_k}=1\right)=\mathbb{P}\left( N_{x_m}=1\right) \mathbb{P}\left(N_{x_k}=0\right).
\end{equation}
\end{assumption}
\begin{assumption}\label{As:Factor}
The point process of the users at the stationary regime $\Phi$ is assumed to be fully characterized by its first two moment measures, where higher moment measures can be factorized into the first two: $\gamma^{(3)}(\cdot,\cdot,\cdot)=\gamma^{(2)}(\cdot,\cdot) \gamma^{(1)}(\cdot)$.
\end{assumption} 
These two assumptions inherently assume some sort of independence between the probability of the presence of a user at different locations within $\mathcal{D}$, and hence, they are not exact. However, we still capture this dependence partially through $\gamma^{(2)}(\cdot,\cdot)$.  Note that even with these two assumptions, we still cannot directly evaluate the expectations in \eqref{Eq:SOH_2} and \eqref{Eq:SOH_8}, since we only have an expression for the mean interference and not its whole distribution. To proceed, we use the following two assumptions.

\begin{assumption}\label{As:FirEq}
Given that a user is located at $x\in\mathcal{D}$ in \eqref{Eq:SOH_2}, the point process of the other users within the cell is assumed to be a PPP with intensity function $\frac{\gamma^{(2)}(x,.)}{\gamma^{(1)}(.)}$.
\end{assumption}
\begin{assumption}\label{As:SecEq}
The interference seen by the users at $x,y \in \mathcal{D}$ in \eqref{Eq:SOH_8}, is assumed to be the average interference given in \eqref{Eq:SOH_6}.
\end{assumption}

Assumption \ref{As:FirEq} is based on matching the mean of the PPP with the mean of the interference found in \eqref{Eq:SOH_6}. The rationale behind it is that since we have seen that the probability to find a user at point $x$ depends on whether or not there are other users in cell, especially users closer to the BS than $x$, then we assume that the statistics of the interference seen by a user at $x$ changes depending on $x$. In other words, at all locations, users suffer from interference stemming from a PPP, but the intensity of this interference depends on the location of the considered user. The same rationale applies to Assumption \ref{As:SecEq}. However, our results show that it is enough only to capture the mean of the interference in \eqref{Eq:SOH_8} which reduces the computation complexity of \eqref{Eq:SOH_8}. 

Both assumptions can have a weak interpretation as a mean-field limit as in the first-order approximation. The difference here is that the mean-field limit depends on the tagged user (the observer). In Assumption 3, the environment is abstracted by the locations of other users in the network, given a user at $x$. In Assumption 4, the environment is abstracted by the interference level in the network given a user at $x$ and a user at $y$. Hence, unlike the first-order approximation, the mean-field limits depend on the location of the observer. Nevertheless, this argument is based on intuition, and it needs to be proven. Hence, we justify these assumptions based on moment matching and delay discussing the connection to mean-field limits in future work.

% {Both assumptions can be justified by the  mean-field limit as in the previous section. }

 The final results in the theorem follow by taking the limit when $\epsilon \rightarrow 0$ of \eqref{Eq:SOH_2} and \eqref{Eq:SOH_8} and then applying the expectation w.r.t. the interference as described by the assumptions \ref{As:FirEq} and \ref{As:SecEq}. Then by applying the assumptions \ref{As:Probs} and \ref{As:Factor} to \eqref{Eq:SOH_8}.

\section{Proof of Theorem 6}\label{App:Thm:GeneralRa}
Define the CTMC $\barbelow{\Phi}^{(d)}$ as in Appendix \ref{App:Thm2} with the exception that the service rate of a user located within $A_i^{(\epsilon)}$ is given by
 \begin{align}\label{Eq:ServRate_gen}
B\mu \log_2 \left(1+ \frac{\barbelow{L}^{(\epsilon)}_i}{\sum\limits_{j=1}^{N_{\epsilon}}k_j\bar{L}^{(\epsilon)}_j+\bS} \right).
\end{align}

Similarly, define the CTMC $\bar{\Phi}^{(d)}$ as in Appendix \ref{App:Thm3} but with the following service rate for a user located within $A_i^{(\epsilon)}$
\begin{align}
B\mu\log_2 \left(1+ \frac{\bar{L}^{(\epsilon)}_i}{(k_i-1)\barbelow{L}^{(\epsilon)}_i+\sum\limits_{\substack{j=1\\j\neq i}}^{N_{\epsilon}}k_j\barbelow{L}^{(\epsilon)}_j+\bS}\right).
\end{align}
		 
		 Based on this, it is straightforward to see that Lemma \ref{Lm:CTMCs} holds in this case also. Moreover, since $\log_2(1+x)\leq \frac{1}{\ln(2)} x, \ \forall x\geq0$, the service rate of a user under the general rate function is less than the service rate of the same user under the service rate given in \eqref{Eq:RateFunction2}. Hence, using stochastic dominance and Theorem \ref{Th:Suf}, we can conclude that the network is unstable for $\lambda>\lambda_c$, where $\lambda_c$ is given by \eqref{Eq:Thm6_1}.
		 
		 To prove that the network is stable for $\lambda<\lambda_c$, we follow the same proof given in Appendix \ref{App:Thm2} with the exception that $q_i$ and $q$ are now given by
		 \begin{align}
		 q_i&=	B\mu k_i \log_2 \left(   \frac{\barbelow{L}^{(\epsilon)}_i}{\sum\limits_{j=1}^{N_{\epsilon}}k_j\bar{L}^{(\epsilon)}_j+\bS}\right),\label{eq:Blah}\\
q&=\sum\limits_{i=1}^{N_{\epsilon}}q_i=	B\mu\sum\limits_{i=1}^{N_{\epsilon}}k_i \log_2 \left( \frac{\barbelow{L}^{(\epsilon)}_i}{\sum\limits_{j=1}^{N_{\epsilon}}k_j\bar{L}^{(\epsilon)}_j+\bS}\right).
\end{align}

Moreover, since we chose $C=\left\{\zeta: \sum\limits_{i=1}^{N_{\epsilon}}k_i\barbelow{L}^{(\epsilon)}_i \leq M\right\}$, where $M \in \mathbb{R}_{+}$, then outside this set, we can lower bound the denominator in \eqref{eq:Blah} by $M$. Hence, by choosing a large enough $M$, the term in \eqref{eq:Blah} can get arbitrarily close to 
\begin{align}
   \frac{B\mu}{\ln(2)}   \frac{ k_i\barbelow{L}^{(\epsilon)}_i}{\sum\limits_{j=1}^{N_{\epsilon}}k_j\bar{L}^{(\epsilon)}_j+\bS}.
\end{align}

Then the proof follows by the same steps in Appendix  \ref{App:Thm2}.
% \section{Summary of Changes}
% 1- Typos.
% 2- Change the channel inversion parameter.
% 3- Typo in Theorem 7.
% 4- Highlight that the result means that if we change the location of the cell, the stability doesn't change.
% 5- Highlight that the analysis can be easily extended to the case where fading exists, but strictly positive and doesn't change during the user's stay in the network.

\bibliographystyle{IEEEtran}
\bibliography{AhmadRef}

\begin{thebibliography}{10}
\providecommand{\url}[1]{#1}
\csname url@samestyle\endcsname
\providecommand{\newblock}{\relax}
\providecommand{\bibinfo}[2]{#2}
\providecommand{\BIBentrySTDinterwordspacing}{\spaceskip=0pt\relax}
\providecommand{\BIBentryALTinterwordstretchfactor}{4}
\providecommand{\BIBentryALTinterwordspacing}{\spaceskip=\fontdimen2\font plus
\BIBentryALTinterwordstretchfactor\fontdimen3\font minus
  \fontdimen4\font\relax}
\providecommand{\BIBforeignlanguage}[2]{{%
\expandafter\ifx\csname l@#1\endcsname\relax
\typeout{** WARNING: IEEEtran.bst: No hyphenation pattern has been}%
\typeout{** loaded for the language `#1'. Using the pattern for}%
\typeout{** the default language instead.}%
\else
\language=\csname l@#1\endcsname
\fi
#2}}
\providecommand{\BIBdecl}{\relax}
\BIBdecl

\bibitem{Stability_AlAmmouri19Conf}
A.~AlAmmouri, J.~G. Andrews, and F.~Baccelli, ``Stability of wireless random
  access systems,'' in \emph{Proc., Allerton Conf. on Comm., Control, and
  Computing}, Sep. 2019, to appear.

\bibitem{The_Atzori10}
L.~Atzori, A.~Iera, and G.~Morabito, ``The internet of things: A survey,''
  \emph{Computer networks}, vol.~54, no.~15, pp. 2787--2805, Apr. 2010.

\bibitem{Data_Bertsekas92}
D.~P. Bertsekas, R.~G. Gallager, and P.~Humblet, \emph{Data networks}.\hskip
  1em plus 0.5em minus 0.4em\relax Prentice-Hall International New Jersey,
  1992, vol.~2.

\bibitem{Ergodicity_Tsybakov79}
B.~S. Tsybakov and V.~A. Mikhailov, ``Ergodicity of a slotted {ALOHA} system,''
  \emph{Problemy peredachi informatsii}, vol.~15, no.~4, pp. 73--87, 1979.

\bibitem{Stability_Szpankowski94}
W.~Szpankowski, ``Stability conditions for some distributed systems: Buffered
  random access systems,'' \emph{Advances in Applied Probability}, vol.~26,
  no.~2, pp. 498--515, Jun. 1994.

\bibitem{Bonald_Wireless04}
T.~Bonald, S.~Borst, N.~Hegde, and A.~Prouti{\'e}re, ``Wireless data
  performance in multi-cell scenarios,'' \emph{SIGMETRICS Perform. Eval. Rev.},
  vol.~32, no.~1, pp. 378--380, Jun. 2004.

\bibitem{On_Rao88}
R.~Rao and A.~Ephremides, ``On the stability of interacting queues in a
  multiple-access system,'' \emph{IEEE Trans. on Info. Theory}, vol.~34, no.~5,
  pp. 918--930, Sep. 1988.

\bibitem{Stability_Borst08}
S.~Borst, M.~Jonckheere, and L.~Leskel{\"a}, ``Stability of parallel queueing
  systems with coupled service rates,'' \emph{Discrete Event Dynamic Systems},
  vol.~18, no.~4, pp. 447--472, Dec. 2008.

\bibitem{Asymptotic_Bordenave12}
C.~Bordenave, D.~McDonald, and A.~Proutiere, ``Asymptotic stability region of
  slotted {Aloha},'' \emph{IEEE Trans. on Info. Theory}, vol.~58, no.~9, pp.
  5841--5855, Sep. 2012.

\bibitem{A_Bordenave07}
C.~Bordenave, D.~R. McDonald, and A.~Prouti{\`e}re, ``A particle system in
  interaction with a rapidly varying environment: Mean field limits and
  applications,'' \emph{Networks \& Heterogeneous Media}, vol.~5, no.~1, pp.
  31--62, Feb. 2010.

\bibitem{Ultimate_Aldous87}
D.~Aldous, ``Ultimate instability of exponential back-off protocol for
  acknowledgment-based transmission control of random access communication
  channels,'' \emph{IEEE Trans. on Info. Theory}, vol.~33, no.~2, pp. 219--223,
  Mar. 1987.

\bibitem{Metastability_Antunes06}
N.~Antunes, C.~Fricker, P.~Robert, and D.~Tibi, ``Metastability of {CDMA}
  cellular systems,'' in \emph{Proc. of the 12th Annual Int. Conf. on Mobile
  Computing and Networking}, ser. MobiCom.\hskip 1em plus 0.5em minus
  0.4em\relax New York, NY, USA: ACM, Sep. 2006, pp. 206--214.

\bibitem{Stochastic_Antunes08}
------, ``Stochastic networks with multiple stable points,'' \emph{The Annals
  of Probability}, pp. 255--278, Jan. 2008.

\bibitem{Toward_Genin08}
D.~Genin and V.~Marbukh, ``Toward understanding of metastability in cellular
  networks: Emergence and implications for performance,'' in \emph{Proc., IEEE
  Globecom}, Nov. 2008, pp. 1--6.

\bibitem{liggett2013stochastic}
T.~M. Liggett, \emph{Stochastic interacting systems: contact, voter and
  exclusion processes}.\hskip 1em plus 0.5em minus 0.4em\relax springer science
  \& Business Media, 2013, vol. 324.

\bibitem{Multi_Vvedenskaya07}
N.~D. Vvedenskaya and Y.~M. Suhov, ``Multi-access system with many users:
  Stability and metastability,'' \emph{Problems of Information Transmission},
  vol.~43, no.~3, pp. 263--269, 2007.

\bibitem{Metastability_Baccelli17}
F.~Baccelli, A.~Rybko, S.~Shlosman, and A.~Vladimirov, ``Metastability of
  queuing networks with mobile servers,'' \emph{Journal of Statistical
  Physics}, pp. 1--25, Nov. 2017.

\bibitem{Stochastic_Baccelli10_2}
F.~Baccelli and B.~B{\l}aszczyszyn, ``Stochastic geometry and wireless
  networks: Volume {II} applications,'' \emph{Foundations and Trends in
  Networking}, vol.~4, no. 1--2, pp. 1--312, 2010.

\bibitem{A_Andrews11}
J.~G. Andrews, F.~Baccelli, and R.~K. Ganti, ``A tractable approach to coverage
  and rate in cellular networks,'' \emph{IEEE Trans. on Communications},
  vol.~59, no.~11, pp. 3122--3134, Nov. 2011.

\bibitem{Stochastic_Haenggi12}
M.~Haenggi, \emph{Stochastic Geometry for Wireless Networks}.\hskip 1em plus
  0.5em minus 0.4em\relax Cambridge University Publishers, 2012.

\bibitem{On_Zhong16}
Y.~Zhong, M.~Haenggi, T.~Q.~S. Quek, and W.~Zhang, ``On the stability of static
  {Poisson} networks under random access,'' \emph{IEEE Trans. on
  Communications}, vol.~64, no.~7, pp. 2985--2998, Jul. 2016.

\bibitem{On_Chisci17}
G.~Chisci, H.~ElSawy, A.~Conti, M.~S. Alouini, and M.~Z. Win, ``On the
  scalability of uncoordinated multiple access for the internet of things,'' in
  \emph{Proc. Int. Symp. Wireless Commun. Systems (ISWCS)}, Aug. 2017, pp.
  402--407.

\bibitem{Spatial_Sankararaman17}
A.~Sankararaman and F.~Baccelli, ``Spatial birth-death wireless networks,''
  \emph{IEEE Trans. on Info. Theory}, vol.~63, no.~6, pp. 3964--3982, Jun.
  2017.

\bibitem{Heterogeneous_Zhong17}
Y.~Zhong, T.~Q.~S. Quek, and X.~Ge, ``Heterogeneous cellular networks with
  spatio-temporal traffic: Delay analysis and scheduling,'' \emph{IEEE Journal
  on Sel. Areas in Communications}, vol.~35, no.~6, pp. 1373--1386, Jun. 2017.

\bibitem{Spatiotemporal_Gharbieh17}
M.~Gharbieh, H.~ElSawy, A.~Bader, and M.~S. Alouini, ``Spatiotemporal
  stochastic modeling of {IoT} enabled cellular networks: {S}calability and
  stability analysis,'' \emph{IEEE Trans. on Communications}, vol.~65, no.~8,
  pp. 3585--3600, Aug. 2017.

\bibitem{Spatiotemporal_Gharbieh18}
M.~{Gharbieh}, H.~{ElSawy}, H.~{Yang}, A.~{Bader}, and M.~{Alouini},
  ``Spatiotemporal model for uplink {IoT} traffic: Scheduling and random access
  paradox,'' \emph{IEEE Trans. on Wireless Communications}, vol.~17, no.~12,
  pp. 8357--8372, Dec. 2018.

\bibitem{Self_Gharbieh19}
M.~{Gharbieh}, A.~{Bader}, H.~{ElSawy}, H.~{Yang}, M.~{Alouini}, and
  A.~{Adinoyi}, ``Self-organized scheduling request for uplink {5G} networks: A
  {D2D} clustering approach,'' \emph{IEEE Trans. on Communications}, vol.~67,
  no.~2, pp. 1197--1209, Feb. 2019.

\bibitem{Performance_Blaszczyszyn15}
B.~B{\l}aszczyszyn, M.~Jovanovic, and M.~K. Karray, ``Performance laws of large
  heterogeneous cellular networks,'' in \emph{Proc., Int. Symp. on Modeling and
  Opt. in Mobile, Ad Hoc, and Wireless Networks (WiOpt)}, May 2015, pp.
  597--604.

\bibitem{A_Crovella98}
M.~E. Crovella, M.~S. Taqqu, and A.~Bestavros, ``Heavy-tailed probability
  distributions in the world wide web,'' \emph{A practical guide to heavy
  tails}, vol.~1, pp. 3--26, Aug. 1998.

\bibitem{A_AlAmmouri19}
A.~{AlAmmouri}, J.~G. {Andrews}, and F.~{Baccelli}, ``A unified asymptotic
  analysis of area spectral efficiency in ultradense cellular networks,''
  \emph{IEEE Trans. on Info. Theory}, vol.~65, no.~2, pp. 1236--1248, Feb.
  2019.

\bibitem{Uplink_Xiao06}
W.~{Xiao}, R.~{Ratasuk}, A.~{Ghosh}, R.~{Love}, Y.~{Sun}, and R.~{Nory},
  ``Uplink power control, interference coordination and resource allocation for
  {3GPP} {E-UTRA},'' in \emph{Proc., IEEE Veh. Technology Conf.}, Sep. 2006,
  pp. 1--5.

\bibitem{Fractional_Jindal08}
N.~Jindal, S.~Weber, and J.~G. Andrews, ``Fractional power control for
  decentralized wireless networks,'' \emph{IEEE Trans. on Wireless
  Communications}, vol.~7, no.~12, pp. 5482--5492, Dec. 2008.

\bibitem{Spatial_Preston75}
C.~Preston, ``Spatial birth and death processes,'' \emph{Advances in applied
  probability}, vol.~7, no.~3, pp. 465--466, Sep. 1975.

\bibitem{A_Parekh93}
A.~K. Parekh and R.~G. Gallager, ``A generalized processor sharing approach to
  flow control in integrated services networks: the single-node case,''
  \emph{IEEE/ACM Trans. on Networking}, vol.~1, no.~3, pp. 344--357, Jun. 1993.

\bibitem{Elements_Baccelli13}
F.~Baccelli and P.~Br{\'e}maud, \emph{Elements of queueing theory: Palm
  Martingale calculus and stochastic recurrences}.\hskip 1em plus 0.5em minus
  0.4em\relax Springer Science \& Business Media, 2013, vol.~26.

\bibitem{Markov_Meyn12}
S.~P. Meyn and R.~L. Tweedie, \emph{Markov chains and stochastic
  stability}.\hskip 1em plus 0.5em minus 0.4em\relax Springer Science \&
  Business Media, 2012.

\bibitem{Markov_Pierre13}
P.~Br{\'e}maud, \emph{Markov chains: Gibbs fields, Monte Carlo simulation, and
  queues}.\hskip 1em plus 0.5em minus 0.4em\relax Springer Science \& Business
  Media, 2013, vol.~31.

\bibitem{Stochastic_Baccelli10}
F.~Baccelli and B.~B{\l}aszczyszyn, ``Stochastic geometry and wireless
  networks: Volume {I} theory,'' \emph{Foundations and Trends in Networking},
  vol.~3, no. 3--4, pp. 249--449, 2010.

\bibitem{Can_Baccelli13}
F.~Baccelli, F.~Mathieu, I.~Norros, and R.~Varloot, ``Can {P2P} networks be
  super-scalable?'' in \emph{Proc., IEEE INFOCOM}, Apr. 2013, pp. 1753--1761.

\bibitem{A_Keilson65}
J.~Keilson, ``A review of transient behavior in regular diffusion and
  birth-death processes. {Part II},'' \emph{Journal of Applied Probability},
  vol.~2, no.~2, pp. 405--428, 1965.

\bibitem{Stochastic_Gallager13}
R.~G. Gallager, \emph{Stochastic processes: theory for applications}.\hskip 1em
  plus 0.5em minus 0.4em\relax Cambridge University Press, 2013.

\bibitem{Topics_Fayolle95}
G.~Fayolle, V.~A. Malyshev, and M.~Menshikov, \emph{Topics in the constructive
  theory of countable Markov chains}.\hskip 1em plus 0.5em minus 0.4em\relax
  Cambridge university press, 1995.

\bibitem{NIST_Olver10}
F.~W. Olver, D.~W. Lozier, R.~F. Boisvert, and C.~W. Clark, \emph{NIST Handbook
  of Mathematical Functions}, 1st~ed.\hskip 1em plus 0.5em minus 0.4em\relax
  New York, NY, USA: Cambridge University Press, 2010.

\bibitem{Handbook_Cuyt09}
A.~A. Cuyt, V.~Petersen, B.~Verdonk, H.~Waadeland, and W.~B. Jones,
  \emph{Handbook of continued fractions for special functions}.\hskip 1em plus
  0.5em minus 0.4em\relax Springer Science \& Business Media, 2008.

\end{thebibliography}
\vfill
\end{document}